\documentclass[a4paper,11pt]{article}

\usepackage{amsmath,amsfonts, amsthm}
\usepackage{bm}
\usepackage{txfonts}
\usepackage{eurosym}
\usepackage{float}
\usepackage{color}
\usepackage{typearea}
\typearea{12}
\usepackage[american]{babel}
\usepackage{csquotes}
\usepackage[style=apa6]{biblatex}
\usepackage{geometry}
\usepackage{caption}
\usepackage[subrefformat=parens]{subcaption}
\usepackage{tabularx}
\usepackage{multirow}

\usepackage{booktabs}
\usepackage{here}
\usepackage[subrefformat=parens]{subcaption}
\usepackage{tabularx}
\usepackage{multirow}

\usepackage{graphicx}

\theoremstyle{definition}
\newtheorem{dfn}{Definition}[section]

\def\diag{\mathop{\rm diag}\nolimits}
\newcommand{\argmax}{\operatornamewithlimits{argmax}}
\newcommand{\argmin}{\operatornamewithlimits{argmin}}

\begin{document}

\title{New approaches of the DCC-GARCH residual: Application to foreign exchange rates}
\author{Kenichiro Shiraya\thanks{Graduate School of Economics, The University of Tokyo. Kenichiro Shiraya is supported by Center for Advanced Research in Finance (CARF).}
\and Kanji Suzuki\thanks{Department of Mathematics, ETH Zurich. Department of Banking and Finance, University of Zurich.}
\and Tomohisa Yamakami\thanks{Graduate School of Economics, The University of Tokyo. Mizuho-DL Financial Technology Co., Ltd. The opinions expressed herein are only those of the authors. They do not represent the official views of the Mizuho-DL Financial Technology Co., Ltd.}}
\date{\today}
\maketitle

\begin{abstract}
Two formulations are proposed to filter out correlations in the residuals of the multivariate GARCH model.
The first approach is to estimate the correlation matrix as a parameter and transform any joint distribution to have an arbitrary correlation matrix.
The second approach transforms time series data into an uncorrelated residual based on the eigenvalue decomposition of a correlation matrix.
The empirical performance of these methods is examined through a prediction task for foreign exchange rates and compared with other methodologies in terms of the out-of-sample likelihood.
By using these approaches, the DCC-GARCH residual can be almost independent.
\end{abstract}

\section{Introduction}

The multivariate GARCH (MGARCH) model has been used to model the time series data of returns on multiple assets (see, e.g., \cite{bauwens2006multivariate}).
The copula-GARCH model, which describes the nonlinear static dependency among assets using copulas,
has also been used in many studies because of its flexibility in modeling dynamic volatility and nonlinear dependency (e.g., \cite{aloui2016relationship}, \cite{kimani2023modelling}). 
However, dependency among assets is generally time varying. Indeed, \cite{ramchand1998volatility} pointed out that the correlation strengthens as the volatility of each return increases in real asset price data.
Dynamic correlation models are studied to address these issues.
The Baba, Engle, Kraft, and Kroner (BEKK) model (\cite{engle1995multivariate}) assumed that conditional variance and covariance interact and vary over time.
Based on these studies, \cite{engle2002dynamic} proposed a parsimonious model called dynamic conditional correlation (DCC).
The DCC model has fewer parameters than the BEKK model, simplifying parameter estimation.
Although the original DCC-GARCH model assumes the normality of the residual,
some studies incorporate higher-order dependencies and dynamic linear correlations in residuals.
For example, \cite{pelagatti2004dynamic} discussed a method to extend the residual to an elliptical distribution.
In addition, we can assume copula-based nonlinearity for the residual in the DCC-GARCH model.

There are two problems with the methods of filtering out correlations in the current literature.
First, the residual of the DCC-GARCH model must be linearly uncorrelated but with higher-order dependency; however, the best distribution to capture it is yet to be concluded.
\cite{kim2016linear} tackled this problem by using the correlation matrix estimated by DCC as the copula parameters and modeling the volatility-filtered return of the GARCH model with time-varying copulas.
\cite{lee2009copula} directly solved the problem by converting a joint distribution into an uncorrelated distribution using linear transformation.
However, numerical calculations are required when the correlation of the distribution cannot be calculated analytically.
Second, the method of filtering out the residual's correlation is not unique.
Depending on the decomposition method, the distributions of the residuals differ.
\cite{troben2007} noted the square root of the matrix and the Cholesky decomposition as specific examples.
However, few studies deal with other variations in decomposition and performance comparisons, as \cite{bauwens2006multivariate} pointed out.

This study proposes two approaches to filter out the correlation between random time series variables.
The first approach is an extension of the method proposed by \cite{lee2009copula}.
Lee and Long filtered out the correlation by linearly transforming the correlation matrix. To do so, they estimated the correlation matrix and computed the square root.
However, this linear transformation cannot be calculated analytically for many distributions and requires numerical calculations.
In this study, we propose a method called the correlation adjustment add-in to avoid numerical calculations. Specifically, the method estimates the transformation matrix as model parameters.
This method can be applied to obtain a distribution with an arbitrary correlation matrix, whereas Lee and Long's approach is limited to obtaining only uncorrelated distributions.
The second approach is a new method for constructing a transformation matrix using eigenvalue decomposition. This method
uses the information in the eigenvalues of the correlation matrix.
This method arises naturally from the principal component analysis (PCA).
Higher-order dependencies among components with small eigenvalues can be regarded as having a small impact on the overall distribution.
Therefore, it may be possible to simplify the model of higher-order dependencies among the components with small eigenvalues.
However, the eigenvalue decomposition does not necessarily ensure that the converted residuals at different points in time follow the same distribution.
This is because eigenvalue decomposition has degrees of freedom for rearranging the basis and for a sign inversion of the basis.
Therefore, we propose a new method to determine the basis order and sign based on the magnitude of the eigenvalues and the inner product of the eigenvectors to achive a close decomposition across time.
In addition to these approaches, we investigate the variation in terms of the decomposition of the correlation matrix or covariance matrix, which has not yet been studied.

In the empirical section, we investigate the performance of our correlation filtering methodologies when applied to the copula-DCC-GARCH model through the exercise to predict foreign exchange rate returns.
This approach is compared with other widely used methodologies, such as the square root of the matrix and Cholesky decomposition.
We also consider the formulation of the transformation matrix for the correlation and covariance matrices.
Notably, a simple copula provides a good out-of-sample fitting if the decomposition method is properly chosen and the correlation adjustment add-in is applied.

The remainder of this paper is organized as follows. 
Section 2 outlines the modeling of the time series data of multi-asset returns using the DCC-GARCH model and the construction of a joint distribution using a copula.
Section 3 describes our new proposed methods.
Section 4 applies the proposed methods to the returns on foreign exchange rates and discusses their effectiveness.
Finally, Section 5 concludes the paper.

\section{Methodology}
The GARCH model represents the typical time series data modeling in the field of financial econometrics.
It is used for forecasting the volatility of asset returns.
To handle multiple assets, a multivariate extension of the GARCH model is used that incorporates a correlation structure.
In this section, we introduce the multivariate GARCH model and the copulas for modeling the residuals.

\subsection{Definitions}
Several symbols used throughout this study are defined.
We assume a probability space $(\Omega,\mathcal{F},\mathbb{P})$ that satisfies the appropriate conditions.
Let $\left\{\mathcal{F}_{t}\right\}_{t\in\{0,1,\cdots,T\}}$ be a filtration of $\mathcal{F}$.
We denote the conditional expectation $\mathcal{F}_{t}$ as $\mathbb{E}_{t}[\cdot]=\mathbb{E}[\cdot|\mathcal{F}_{t}]$.
Let $\bm{x}:=(x_{1},\cdots,x_{N})'$ be a vector of $\mathbb{R}^{N}$.
We use the notation that a function with a vector argument is identical to an $N$-variable function, as $f(\bm{x})=f(x_{1},\cdots, x_{N})$.
A diagonal matrix with $x_{1},\cdots,x_{N}$ as its diagonal component is denoted by $\text{diag}(\bm{x}):=\text{diag}(x_{1},\cdots,x_{N})$.
Let $Q$ be a square matrix and $\text{diag}\{Q\}$ be a diagonal matrix consisting of only the diagonal elements of $Q$.
Let $\partial_{\bm{x}}$ be the differential operator as $\frac{\partial^{N}}{\partial x_{1}\cdots\partial x_{N}}$.
For a semi-positive definite matrix $A$, we write $\sqrt{A}$ for the square root of matrix $A$,
where $\sqrt{A}$ satisfies $A=\left(\sqrt{A}\right)^{2}$ and is a semi-positive definite matrix.

\subsection{DCC-GARCH Model}
\label{section:DCC-GARCH}
We also consider the rate of return on $N$ assets. 
Let $\bm{r}_{t}:=(r_{1,t},\cdots,r_{N,t})'$ be the rate of return observed at time $t$.
Let $\bm{\xi}_{t}:=(\xi_{1,t},\cdots,\xi_{N,t})'$ be the noise with a mean of 0 and a variance of 1 under $\mathcal{F}_{t-1}$ and $\bm{\sigma}_{t}:=(\sigma_{1,t},\cdots,\sigma_{N,t})'$ be the volatility.
We assume that the $i$-th rate of return is described as
\begin{eqnarray}
r_{i,t}:=\sigma_{i,t}\xi_{i,t}.\label{formula:logreturn}
\end{eqnarray}
Focusing on the $i$-th asset, the GARCH(1,1) model is defined as
\begin{eqnarray}
  \sigma_{i,t}^{2}:=\omega_{i}+\alpha_{i} r_{i,t-1}^{2} + \beta_{i} \sigma_{i,t-1}^{2},\label{formula:GARCH}
\end{eqnarray}
where $\omega_{i}>0$, $\alpha_{i}\geq0$, $\beta_{i}\geq0$ and $\alpha_{i}+\beta_{i}<1$ are satisfied. These restrictions ensure the non-negativity and stationarity of the process.
Here, $\mathbb{E}[r_{i,t}^{2}]=\mathbb{E}[\sigma_{i,t}^{2}]$ holds.
Furthermore, by using the stationarity assumption ($\bar{\sigma}_{i}^{2}:=\mathbb{E}[r_{i,t}^{2}]=\mathbb{E}[r_{i,t-1}^{2}]$), we obtain
\begin{eqnarray}
  \mathbb{E}[r_{i,t}^{2}]=\frac{\omega_{i}}{1-\alpha_{i}-\beta_{i}}
\end{eqnarray}
by using the expected value of Equation \eqref{formula:GARCH}.
For time series data $\bm{r}_{t}$, we call the residual $\bm{\xi}_{t}$ filtered by the GARCH model as the GARCH residual.

We assume the correlation structure for $\bm{\xi}_{t}$.
In the DCC model proposed by \cite{engle2002dynamic}, conditional correlation varies with time.
This implies that the $R_{t}$ is defined as the correlation matrix of $\bm{r}_{t}$ under $\mathcal{F}_{t-1}$ as
\begin{eqnarray}
  R_{t}:=\mathbb{E}_{t-1}\left[\bm{\xi}_{t}\bm{\xi}'_{t}\right].
\end{eqnarray}
The DCC model assumes the dynamics in the correlation matrix under the stationarity conditions of $a\geq0, b\geq0, a+b<1$ as follows:
\begin{eqnarray}
  Q_{t}&:=&(1-a-b)\bar{Q}+a\bm{\xi}_{t}\bm{\xi}'_{t} +bQ_{t-1},\label{formula:DCC1}\\
  R_{t}&=&\sqrt{\text{diag}\{Q_{t}\}}^{-1}Q_{t}\sqrt{\text{diag}\{Q_{t}\}}^{-1}
\end{eqnarray}
where $\bar{Q}:=\mathbb{E}[\bm{\xi}_{t}\bm{\xi}'_{t}]$ and $\bar{Q}$ is a matrix with diagonal components of 1.
From the stationarity assumption for $Q_{t}$ ($\mathbb{E}[Q_{t}]=\mathbb{E}[Q_{t-1}]$),
we obtain $\mathbb{E}[Q_{t}]=\bar{Q}$ by taking the expected value of Equation \eqref{formula:DCC1}.

Furthermore, let $\Xi_{R_{t}}$ be a decomposition satisfying $R_{t}=\Xi_{R_{t}}\Xi'_{R_{t}}$.
We model $\bm{\xi}_{t}$ with a linearly uncorrelated random vector $\bm{\epsilon}_{t}:=(\epsilon_{1,t},\cdots,\epsilon_{N,t})'$ as
\begin{eqnarray}
  \bm{\xi}_{t}=\Xi_{R_{t}}\bm{\epsilon}_{t}.\label{formula:residual}
\end{eqnarray}
We emphasize that a linearly uncorrelated vector is not necessarily independent.
If the residual $\bm{\epsilon}_{t}$ follows an independent standard normal distribution,
the model is identical for any given decomposition owing to its reproductive properties.
However, if we assume a general distribution as $\bm{\epsilon}_{t}$, the distribution of $\bm{\xi}_{t}$ depends on the decomposition of $\Xi_{R_{t}}$.
We propose a new decomposition method in Section \ref{section:innovation} and compare our method with several alternatives in Section \ref{section:EmpiricalPerformance}.
Some studies formulate the return as $r_{t}=\Xi_{H_{t}}\epsilon_{t}$, with the covariance matrix
$H_{t}:=\diag(\bm{\sigma}_{t})R_{t}\diag(\bm{\sigma}_{t})$ and its decomposition $H_{t}=\Xi_{H_{t}}\Xi'_{H_{t}}$.
This decomposition using $\Xi_{H_{t}}$ is nested in our formulation if we define $\Xi_{R_{t}}=\diag(\bm{\sigma}_{t})^{-1}\Xi_{H_{t}}$.
The methods proposed in this study are not restricted to applying the DCC-GARCH model but also to the multivariate GARCH model, whose return is formulated by $r_{t}=\Xi_{\tilde{H}_{t}}\epsilon_{t}$,
where $\tilde{H}_{t}$ denotes the covariance matrix.
However, for simplicity, the DCC-GARCH model is assumed hereafter.
The residual $\bm{\epsilon}_{t}$ filtered by the DCC model is called the DCC residual.

In this study, we consider a model in which $\bm{\epsilon}_{t}$ follows a time-invariant joint distribution $F_{D}(\cdot)$, and $F_{D}(\cdot)$ is constructed using copulas (copula-DCC-GARCH model).
For comparison in Section \ref{section:EmpiricalPerformance}, we consider the copula-GARCH model, which is a static correlation model, assuming that $\bm{\xi}_{t}$ follows a joint distribution $F_{C}(\cdot)$ independent of $t$,
 and $F_{C}(\cdot)$ is expressed using the marginal distribution of $\bm{\xi}_{t}$ and the copulas.

\subsection{Copula}
This section outlines the general properties of the copulas and several specific copulas.
For example, see \cite{jaworski2010copula} for more information on the various properties of copulas.
\subsubsection{Copula Properties}
\label{section:CopulaProperties}
A copula is a function that links marginal and joint distributions.
In practice, it is used to construct a joint distribution from marginal distributions.
\begin{dfn}
The copula $C(\bm{u})$ is a function of $[0,1]^{N}\to[0,1]$ with the following properties:
\begin{itemize}
  \item When at least one element of $\bm{u}=(u_{1},\cdots,u_{N})$ is 0, $C(\bm{u})=0$ is satisfied.
  \item For any element $u_{i}$ of $\bm{u}$, $C(\bm{u})=u_{i}$ is satisfied when any value other than $u_{i}$ is 1.
  \item For any hypercube $[u_{1,l},u_{1,u}]\times\cdots\times[u_{N,l},u_{N,u}]\subseteq[0,1]^{N}$, 
  \begin{eqnarray}
    \Delta_{u_{N,l}}^{u_{N,u}}\cdots\Delta_{u_{1,l}}^{u_{1,u}}C(\bm{u})\geq0
  \end{eqnarray}
holds, where $\Delta_{u_{i,l}}^{u_{i,u}}C(\bm{u})=C(u_{1},\cdots,u_{i-1},u_{i,u},u_{i+1},\cdots,u_{N})-C(u_{1},\cdots,u_{i-1},u_{i,l},u_{i+1},\cdots,u_{N})$. 
\end{itemize}
\end{dfn}

When an $N$-dimensional joint distribution $F(\bm{x})=F(x_{1},\cdots,x_{N})$ has marginal distributions $F_{1}(x_{1}),\cdots,F_{N}(x_{N})$, 
there exists an $N$-dimensional copula $C_{F}(\bm{u})$ related as
\begin{eqnarray}
  F(\bm{x})=C_{F}(F_{1}(x_{1}),\cdots,F_{N}(x_{N})).
\end{eqnarray}
In particular, copula $C_{F}(\bm{u})$ is uniquely determined when the marginal distributions are continuous.
Moreover, when the marginal distributions have inverse functions, they can be written as
\begin{eqnarray}
  C_{F}(\bm{u})=F(F_{1}^{-1}(u_{1}),\cdots,F_{N}^{-1}(u_{N})).
\end{eqnarray}
We call this the copula derived from the joint distribution $F$.
Additionally, we assume that the joint density function of $F(\bm{x})$ is $f(\bm{x})$,
and $f_{1},\cdots,f_{N}$ are the marginal density functions of $F_{1},\cdots,F_{N}$, which satisfies
\begin{eqnarray}
  f(\bm{x})=\partial_{u}C_{F}(F_{1}(x_{1}),\cdots,F_{N}(x_{N}))\prod_{i=1}^{N}f_{i}(x_{i})
\end{eqnarray}
where $\partial_{u}C_{F}(\bm{u})$ is the copula density.

For a two-dimensional copula, we consider copulas with their densities rotated at $90^{\circ}$, $180^{\circ}$, and $270^{\circ}$ in a two-dimensional plane.
Let $C(u,v)$ be the original copula.
The $90^{\circ}$-rotated copula is $C_{90^{\circ}}(u,v):=u-C(1-v,u)$.
The $180^{\circ}$-rotated copula is $C_{180^{\circ}}(u,v):=C(1-u,1-v)+u+v-1$. 
The $270^{\circ}$-rotated copula is $C_{270^{\circ}}(u,v):=v-C(v,1-u)$.

\subsubsection{Parametric Copulas}
In this section, we explain the major copulas used.
\begin{itemize}
  \item Archimedean copula\\
  Archimedean copulas are a family of copulas parameterized by the generator function $\phi$.
  These copulas are described as 
  \begin{eqnarray}
    \mathcal{C}(u_{1},\cdots,u_{N})=\phi^{-1}(\phi(u_{1})+\cdots+\phi(u_{N})).
  \end{eqnarray}
  The following Archimedean copula is often used: 
  \begin{eqnarray}
    \phi_{cr}(t)&:=&\frac{1}{\theta}(t^{-\theta}-1)\hspace{5mm}(\mbox{Crayton copula})\\
    \phi_{fr}(t)&:=&\log\left(\frac{e^{\theta u}-1}{e^{\theta}-1}\right)\hspace{5mm}(\mbox{Frank copula})\\
    \phi_{gu}(t)&:=&(-\log(u))^{\theta}\hspace{5mm}(\mbox{Gumbel copula})
  \end{eqnarray}
As the same expression is obtained even if the arguments are swapped,
the asymmetrical correlation structure in terms of argument swapping cannot be expressed by the Archimedean copula.
  \item Plackett copula\\
The copula expressed in the following equation is called a two-dimensional Plackett copula.
    \begin{eqnarray}
      C_{Pl(\theta)}:=\frac{[1+(\theta-1)(u_{1}+u_{2})]-\sqrt{[1+(\theta-1)(u_{1}+u_{2})]^{2}-4u_{1}u_{2}\theta(\theta-1)}}{2(\theta-1)},
    \end{eqnarray}
  \item Gaussian copula\\
The Gaussian copula is derived from the multivariate standard normal distribution $\Phi_{\Sigma}(\bm{x})$ with the correlation matrix $\Sigma$
and the standard normal distribution function $\Phi(x)$ as
    \begin{eqnarray}
      C_{\Phi_{\Sigma}}(u_{1},\cdots,u_{N}):=\Phi_{\Sigma}\left(\Phi^{-1}(u_{1}),\cdots,\Phi^{-1}(u_{N})\right).
    \end{eqnarray}
    \item $t$ copula\\
Let $t_{\Sigma,\nu}(\bm{x})$ be a multivariate $t$ distribution, with parameters as a positive definite matrix $\Sigma$ of all diagonal components 1 and the degree of freedom $\nu$,
and $t_{\nu}(x)$ as a univariate $t$ distribution with the degree of freedom $\nu$.
$t$ copula is derived from a multivariate $t$ distribution as
    \begin{eqnarray}
      C_{t_{\Sigma,\nu}}(u_{1},\cdots,u_{N}):=t_{\Sigma,\nu}\left(t_{\nu}^{-1}(u_{1}),\cdots,t_{\nu}^{-1}(u_{N})\right).
    \end{eqnarray}
\end{itemize}

\subsubsection{Pair Copula}
A pair copula is a method for constructing a multidimensional copula by combining two-dimensional copulas.
Let $(X_{1},\cdots,X_{N})$ be a random variable vector, $F_{1}(x_{1}),\cdots,F_{N}(x_{N})$ and $f_{1}(x_{1}),\cdots,f_{N}(x_{N})$
be the univariate distributions and densities of $X_{1},\cdots,X_{N}$.
We also define $F_{ij}(x_{i},x_{j})$ as the joint distribution and $f_{ij}(x_{i},x_{j})$ as the joint density of $X_{i}$ and $X_{j}$ with $i,j\in \{1, \cdots, N\}, i\neq j$.
$C_{F_{ij}}(u_{i},u_{j}), c_{F_{ij}}(u_{i},u_{j})$ are the copula and copula densities derived from $F_{ij}$. 
The following relationships hold between these functions:
\begin{eqnarray}
  F_{ij}(x_{i},x_{j})&=&C_{F_{ij}}(F_{i}(x_{i}),F_{j}(x_{j})),\\
  f_{ij}(x_{i},x_{j})&=&c_{F_{ij}}(F_{i}(x_{i}),F_{j}(x_{j}))f_{i}(x_{i})f_{j}(x_{j}),\\
  F_{ij}(x_{i},x_{j})&=&F_{ji}(x_{j},x_{i}).
\end{eqnarray}
Additionally, the distribution function of $X_{i}$ under the condition $X_{j}$ can be written as
\begin{eqnarray}
  F_{i|j}(x_{i}|x_{j})&:=&\int_{-\infty}^{x_{i}}f_{i|j}(x|x_{j})dx=\left.\frac{\partial}{\partial u_{j}}C_{ij}(F_{i}(x_{i}),u_{j})\right|_{u_{j}=F_{j}(x_{j})},\\
  f_{i|j}(x_{i}|x_{j})&:=&\frac{f_{ij}(x_{i},x_{j})}{f_{j}(x_{j})}=c_{ij}(F_{i}(x_{i}),F_{j}(x_{j}))f_{i}(x_{i}).
\end{eqnarray}
Furthermore, the probability density and distribution of the $X_{k}$
conditional on $X_{i_{1}},\cdots,X_{i_{m}}$, where $i_{1},\cdots,i_{m} (\neq k)$ and $m\geq2$, can be written as
\begin{eqnarray}
  f_{k|i_{1},\cdots,i_{m}}(x_{k}|x_{i_{1}},\cdots,x_{i_{m}})&:=&\frac{f_{i_{m}k|i_{1},\cdots,i_{m-1}}
  (x_{i_{m}},x_{k}|x_{i_{1}},\cdots,x_{i_{m-1}})}{f_{i_{m}|i_{1},\cdots,i_{m-1}}(x_{i_{m}}|x_{i_{1}},\cdots,x_{i_{m-1}})}\nonumber\\
  &=&c_{i_{m}k|i_{1},\cdots,i_{m-1}}(F_{i_{m}|i_{1},\cdots,i_{m-1}}(x_{i_{m}}|x_{i_{1}},\cdots,x_{i_{m-1}}),F_{k|i_{1},\cdots,i_{m-1}}(x_{k}|x_{i_{1}},\cdots,x_{i_{m-1}}))\nonumber\\
  &&\times f_{k|i_{1},\cdots,i_{m-1}}(x_{k}|x_{i_{1}},\cdots,x_{i_{m-1}}), \label{formula:PairCopulaBlock}\\
  F_{k|i_{1},\cdots,i_{m}}(x_{k}|x_{i_{1}},\cdots,x_{i_{m}})&:=&\int_{-\infty}^{x_{k}}f_{k|i_{1},\cdots,i_{m}}(x|x_{i_{1}},\cdots,x_{i_{m}})dx,
\end{eqnarray}
where $c_{i_{m}k|i_{1},\cdots,i_{m-1}}$ is the copula density of the relation $X_{i_{m}},X_{k}$ under the condition $X_{i_{1}},\cdots,X_{i_{m-1}}$. 
The joint density functions for $X_{1},\cdots,X_{N}$ are constructed arbitrarily in order of the subscriptions.
For example, 
\begin{eqnarray}
  f_{1,\cdots,N}&=&f_{N}f_{N-1|N}f_{N-2|N-1,N}\cdots f_{1|2,\cdots,N}
\end{eqnarray}
is one expression. 
If \eqref{formula:PairCopulaBlock} is applied recursively, it can be expressed using a two-variable copula.
Particularly in the case of three variables, there are three expressions as follows:
\begin{eqnarray}
  f_{123}
  &=&c_{23|1}(F_{2|1}(x_{2}|x_{1}),F_{3|1}(x_{3}|x_{1}))c_{13}(F_{1}(x_{1}),F_{3}(x_{3}))c_{12}(F_{1}(x_{1}),F_{2}(x_{2}))\nonumber\\
  &&\cdot f_{1}(x_{1})f_{2}(x_{2})f_{3}(x_{3}),\\
  &=&c_{13|2}(F_{1|2}(x_{1}|x_{2}),F_{3|2}(x_{3}|x_{2}))c_{12}(F_{1}(x_{1}),F_{2}(x_{2}))c_{23}(F_{2}(x_{2}),F_{3}(x_{3}))\nonumber\\
  &&\cdot f_{1}(x_{1})f_{2}(x_{2})f_{3}(x_{3}),\\
  &=&c_{12|3}(F_{1|3}(x_{1}|x_{3}),F_{2|3}(x_{2}|x_{3}))c_{13}(F_{1}(x_{1}),F_{3}(x_{3}))c_{23}(F_{2}(x_{2}),F_{3}(x_{3}))\nonumber\\
  &&\cdot f_{1}(x_{1})f_{2}(x_{2})f_{3}(x_{3}),
\end{eqnarray}
We call these Pivot1, Pivot2, and Pivot3 in this study.

\subsection{Parameter Estimation}
We employ a three-step estimation for the copula-DCC-GARCH model.
In the first and second steps, quasi-maximum likelihood estimation (QMLE) is used to estimate the parameters of the DCC-GARCH model.
Let $\theta_{i}:=(\bm{\omega}_{i}, \bm{\alpha}_{i}, \bm{\beta}_{i}, \sigma_{i,0})$ be the parameters of the GARCH model
of the $i$-th asset and $\bm{\theta}:=(\theta_{1},\cdots,\theta_{N})$ be the collection on them.
Let $\psi:=(a,b,\bar{Q}, Q_{0})$ be the parameters of DCC,
and let $\pi_{j} (j=2, 3)$ be the parameters of the joint distribution $p_{\pi_{j}}(\bm{x})$ of $N$ variables. We use $p_{\pi_{j}}(\bm{x})$ for the residual distribution.
The quasi-log-likelihood function $LL(\bm{\theta},\psi)$ of $\bm{\xi}_{t}$ can be written as:
\begin{eqnarray}
  LL(\bm{\theta},\psi)&:=&LL_{V}(\bm{\theta})+LL_{C}(\bm{\theta},\psi),\\
  LL_{V}(\bm{\theta})&:=&-\frac{1}{2}\sum_{i=1}^{N}\sum_{t}\left(\log(2\pi)+\log(\sigma_{i,t}^{2})+\frac{r_{i,t}^{2}}{\sigma_{i}^{2}}\right),\\
  LL_{C}(\bm{\theta},\psi)&:=&-\frac{1}{2}\sum_{t}\left(\log|R_{t}|+\bm{\xi}'_{t}R_{t}\bm{\xi}_{t}-\bm{\xi}'_{t}\bm{\xi}_{t}\right).
\end{eqnarray}
Here, $LL_{V}(\bm{\theta})$ is the sum of the GARCH model's log-likelihood functions for all assets, assuming that the residual $\xi_{t,i}$ follows a normal distribution.
This implies that the result is the same as that of the quasi-maximum likelihood estimation for each asset. 
\cite{engle2002dynamic} proposed that 
the parameters of the GARCH model can be determined by optimizing
\begin{eqnarray}
  \hat{\bm{\theta}}:=\argmax_{\bm{\theta}}\{LL_{V}(\bm{\theta})\},
\end{eqnarray}
as the first step. Furthermore, this can be performed independently for each asset.

In the second step, we optimize
\begin{eqnarray}
  \hat{\psi}:=\argmax_{\psi}\{LL_{C}(\hat{\theta},\psi)\}.\label{formula:DCCCalibration}
\end{eqnarray}

In the third step, the DCC residual $\bm{\hat{\epsilon}}_{t}:=\Xi_{\hat{R}_{t}}^{-1}\hat{\bm{\xi}}_{t}$ is calculated
using the GARCH residual $\hat{\bm{\xi}}_{t}$ and correlation matrix $\hat{R}_{t}$ given the estimates in the first and second steps.
We obtain the parameters $\hat{\pi}_{3}$ from the maximum likelihood estimation of the joint distribution using this residual.
In mathematical terms, the log-likelihood function is expressed as follows:
\begin{eqnarray}
  LL_{D3}(\pi_{3}):=\sum_{t}\log\left(p_{\pi_{3}}(\hat{\epsilon}_{t})\right),
\end{eqnarray}
and we calculate
\begin{eqnarray}
  \hat{\pi}_{3}:=\argmax_{\pi_{3}}\left\{LL_{D3}(\pi_{3})\right\}.
\end{eqnarray}
For the copula-GARCH model, we skip the estimation of DCC parameters and conduct a two-step estimation,
which implies that the objective function and estimates are
\begin{eqnarray}
  LL_{D2}(\pi_{2})&:=&\sum_{t}\log\left(p_{\pi_{2}}(\hat{\xi}_{t})\right),\\
  \hat{\pi}_{2}&:=&\argmax_{\pi_{2}}\left\{LL_{D2}(\pi_{2})\right\}.
\end{eqnarray}

\section{Innovation}
\label{section:innovation}
There are two issues with extending the DCC-GARCH model to non-Gaussian residuals.
First, a linearly uncorrelated distribution with a higher-order dependency is required for modeling the distribution of DCC residual $\bm{\epsilon}_{t}$; however, distributions with such properties are not widely known.
Second, the decomposition $\Xi_{R_{t}}$ is not unique and has not been fully studied.
\cite{lee2009copula} proposed a method to uncorrelate the copula-based distribution by applying a linear transformation.
However, numerical calculations are required if a linear correlation cannot be obtained analytically.
They also use the square root of the matrix as the decomposition $\Xi_{R_{t}}$, but there is still room to look for alternatives.

Our study proposes original methods for two aspects of the DCC-GARCH framework.
First, we propose a correlation adjustment add-in to transform a joint distribution into a distribution with an arbitrary correlation matrix that fits the empirical correlation.
This is useful when analytical computations are not feasible.
We apply the transformed distribution to the DCC residual $\bm{\epsilon}_{t}$ and the GARCH residual $\bm{\xi}_{t}$.
Second, we propose a new method for constructing $\Xi_{R_{t}}$ using eigenvalue decomposition.
This is determined from the time series of the eigenvectors to achieve the decomposition with the highest similarity in direction.
In addition to these original methodologies, we investigate the differences in decomposition 
$R_{t}=\Xi_{R_{t}}\Xi'_{R_{t}}$. These include the square root of the matrix, the Cholesky decomposition,
and decomposition based on the correlation matrix or covariance matrix,
as this variation has not been sufficiently investigated.

\subsection{Correlation Adjustment Add-in}
In the first approach, we introduce a correlation adjustment add-in, which is an extension of \cite{lee2009copula}.
They used the following transformation to obtain an uncorrelated distribution:
\\

\textbf{Lee and Long's method:}
Let $F_{X}$ be the joint distribution followed by an $N$ random variable vector $X$ with mean $\bm{0}$; let $S_{X}$ be its covariance matrix.
$\left(\sqrt{S_{X}}\right)^{-1}X$ follows a linear uncorrelated joint distribution with a mean of $0$ and a variance of $1$.
Under the DCC-GARCH model, the covariance matrix of the rate of return is defined as $H_{t}$,
and the rate of return is written as $\bm{r}_{t}=\sqrt{H_{t}}\bm{\hat{\epsilon}}_{t}$, where the residual follows $\bm{\hat{\epsilon}}_{t}\sim \left(\sqrt{S_{X}}\right)^{-1}X$.
\\

This method requires computing covariance matrix $S_{X}$ from joint distribution $F_{X}$.
If not obtained analytically, the computational burden may increase because of the curse of dimensionality.
To avoid this problem, our approach estimates $S_{X}$ and additional correlation adjustment as parameters.
Our methodologies can be applied to the copula-GARCH case as well as the DCC residual and the copula-GARCH case because the methodologies allow for correlated distributions.
\\

\textbf{Correlation Adjustment Add-in:}
Let $\bm{Y}=(Y_{1},\cdots,Y_{N})$ be an $N$-dimensional random variable following the joint distribution $F_{\bm{Y}}$, with a mean of 0 and a variance of 1, and let $S_{\bm{Y}}$ be its correlation matrix.
We then consider the transformation of $Y$ into $N$-dimensional random variables with covariance matrix $J$, whose $(1,1)$ element is $1$.
$J$ represents the covariance matrix of the converted distribution after the correlation adjustment add-in.
We define $L_{J}$ and $L_{S_{\bm{Y}}}$ as Cholesky decompositions of $J$ and $S_{\bm{Y}}$ that satisfy
$J=L_{J}L'_{J}$ and $S_{Y}=L_{S_{\bm{Y}}}L'_{S_{\bm{Y}}}$.
Since the upper-left element of the Cholesky decomposition of the correlation matrix is $1$,
the product of $L_{J}$ and $L_{S_{Y}}^{-1}$ is a lower triangular matrix of the form
\begin{eqnarray}
  L_{J,S_{Y}}:=L_{J}L_{S_{Y}}^{-1}=\begin{pmatrix}
  1&&&0\\
  a_{2,1}&a_{2,2}&&\\
  \vdots&&\ddots&\\
  a_{N,0}&\hdots&&a_{N,N}
\end{pmatrix}.
\end{eqnarray}
Since the covariance matrix of $L_{J,S_{Y}}Y$ is $J$,
the joint distribution of $L_{J,S_{Y}}Y$ can have any correlation matrix depending on the parameters $a_{2,1},\cdots,a_{N,N}$.
\\

Using this formulation, we propose to model the DCC residual as $\bm{\epsilon}_{t}=L_{J,S_{Y}}Y$ and estimate the elements of $L_{J,S_{Y}}$ as parameters.
$L_{J,S_{Y}}Y$ can represent not only the uncorrelated case but also the random variables following any correlation matrix specified by $J$.
In our applications, we use an uncorrelated case for modeling DCC residual $\bm{\epsilon}_{t}$.
We use the correlated case for GARCH residual $\bm{\xi}_{t}$, which provides the static dependency of the Copula-GARCH model.

\subsection{Correlation Matrix Decomposition}
\label{section:correlation_matrix_decomposition}
We introduce the second approach as the correlation matrix decomposition using eigenvalue decomposition.
\cite{engle2002dynamic} presumed the residual to be normal
to ensure that variations in decomposition $\Xi_{R_{t}}$ of $R_{t}$ do not affect the model.
Conversely, when the non-normality of the residual is assumed, the residual distribution varies based on the decomposition method.
However, no studies have compared the differences in decomposition methods.
The square root of a matrix and the Cholesky decomposition are well-known methods in which the decomposition is unique.
The construction of $\Xi_{R_{t}}$ by using eigenvalue decomposition follows naturally from the PCA concept.
However, an algorithm for selecting a specific value is required because eigenvalue decomposition is not unique.
No studies have found the construct $\Xi_{R_{t}}$ by using eigenvalue decomposition.

In this study, we propose a new method to uniquely determine the eigenvalue decomposition for time series matrices.
Specifically, the proposed method minimizes the distance of eigenvectors over time.
\\

\textbf{Time Series Eigenvectors Sort:}
Let $R_{t}=V_{t}D_{t}V'_{t}$ be the eigenvalue decomposition of correlation matrix $R_{t}$.
$\mbox{Eig}_{R_{t}}=V_{t}\sqrt{D_{t}}$ satisfies $R_{t}=\mbox{Eig}_{R_{t}}\mbox{Eig}'_{R_{t}}$.
The uncorrelated residual transformed by $\mbox{Eig}_{R_{t}}$ corresponds to the PCA score.
However, because eigenvalue decomposition entails arbitrariness in rearranging the basis and its sign inversion,
we cannot simply compare the scores between the different time points.
In our approach, assuming that the similarity of basis vectors is high between nearby points in time,
the optimal decomposition series of $R_{t}$ is determined using the levels of angles formed by the eigenvectors.

Let $d_{1,t},\cdots,d_{N,t}$ denote the sequence of eigenvalues of $R_{t}$ in descending order (cases where the eigenvalues are equal are ignored because they rarely occur in reality).
Let $v_{1,t},\cdots,v_{N,t}$ denote the corresponding eigenvectors.
For $t=1$, let $V_{1}^{*}:=(v_{1,1},\cdots,v_{N,1}), D_{1}^{*}:=\diag(d_{1,1},\cdots,d_{N,1})$ and decompose $R_{1}$ as $\mbox{Eig}_{R_{1}}^{*}:=V_{1}^{*}\sqrt{D_{1}^{*}}$.
For $t>1$, the sign that minimizes the sum of squares of the angles formed with each eigenvector in the past $\tau$ period is selected as the eigenvector (In Section \ref{section:ResidualFitting}, we set $\tau=50$.)
Let $d_{c}(\bm{u},\bm{v})$ be the angle formed by the vectors $\bm{u},\bm{v}$
and $\bm{s}_{t}=(s_{1,t},\cdots,s_{N,t})'\in \{-1, 1\}^{N}$.
We define the decomposition as follows:
\begin{eqnarray}
\bm{s}^{*}_{t}&:=&(s_{1,t}^{*},\cdots,s_{N,t}^{*})':=\argmin_{\bm{s}_{t}}\sum_{k=1}^{\min(t,\tau)}\sum_{i=1}^{N}d_{c}^{2}(s_{i,t}v_{i,t},s_{i,t-k}v_{i,t-k}^{*}),\label{formula:EigenDecompMethodMax}\\
V_{t}^{*}&:=&(v_{1,t}^{*},\cdots,v_{N,t}^{*}):=(s_{1,t}^{*}v_{1,t},\cdots,s_{N,t}^{*}v_{N,t}),\\
D_{t}^{*}&:=&\diag(d_{1,t},\cdots,d_{N,t}),\\
\Xi_{t}&=&\mbox{Eig}_{R_{t}}^{*}:=V_{t}^{*}\sqrt{D_{t}^{*}}.\label{formula:EigenDecompMethod}
\end{eqnarray}
We describe the conditions under which this decomposition is not unique.
Since the expression \eqref{formula:EigenDecompMethodMax} can be minimized independently for each dimension $i$,
the condition that $s_{i,t}^{*}$ takes both $1$ and $-1$ as solutions is that
\begin{eqnarray}
  \sum_{k=1}^{\min(t,\tau)}d_{c}^{2}(v_{i,t},s_{i,t-k}v_{i,t-k}^{*})&=&\sum_{k=1}^{\min(t,\tau)}\left(\pi-d_{c}(v_{i,t},s_{i,t-k}v_{i,t-k}^{*})\right)^{2}
\end{eqnarray}
for given $s_{i,u}$ up to $u<t$. Expanding the equation yields
\begin{eqnarray}
\sum_{k=1}^{\min(t,\tau)}d_{c}(v_{i,t},s_{i,t-k}v_{i,t-k}^{*})&=&\frac{\pi}{2}\min(t,\tau).
\end{eqnarray}
It is rare for the angles formed by the eigenvalues to satisfy this condition exactly; therefore, it can be ignored in applications.
\\

Another possible variation is the decomposition of the correlation or covariance matrices.
However, these differences have not been compared in previous studies.
Consequently, we define and compare the following variations in the decompositions:

\begin{itemize}
\item (Sqrt) The square root of the correlation matrix\\
We use $\Xi_{R_{t}}=\sqrt{R_{t}}$,
that is, let $R_{t}=V_{t}D_{t}V'_{t}$ be the eigenvalue decomposition of $R_{t}$, and compute
the matrix as $\sqrt{R_{t}}=V_{t}\sqrt{D_{t}}V'_{t}$.
This decomposition is uniquely determined by the property of the square root of the matrix.

\item (Sqrt2) The square root of the covariance matrix\\
We use the square root of the matrix to the covariance matrix $H_{t}:=\diag(\bm{\sigma}_{t})R_{t}\diag(\bm{\sigma}_{t})$.
The formula is $\Xi_{R_{t}}=\diag(\bm{\sigma}_{t})^{-1}\sqrt{H_{t}}$, and this decomposition is uniquely determined.

\item (Cholesky) Cholesky decomposition\\
The Cholesky decomposition $L_{A}$ is a lower triangular matrix with a positive diagonal satisfying $A=L_{A}L'_{A}$
for a positive definite matrix $A$, which is unique.
We use $\Xi_{R_{t}}=L_{R_{t}}$ as the correlation matrix decomposition.
As $L_{R_{t}}=\diag(\bm{\sigma}_{t})^{-1}L_{H_{t}}$ is satisfied,
this decomposition is the same as that of Cholesky decomposition's covariance matrix.

\item (Eigen) Eigenvalue decomposition of a correlation matrix\\
We apply the formula \eqref{formula:EigenDecompMethod}.

\item (Eigen2) Eigenvalue decomposition of a covariance matrix\\
We apply the (Eigen) decomposition to covariance matrix $H_{t}$, that is, $\Xi_{R_{t}}=\diag(\bm{\sigma}_{t})^{-1}\mbox{Eig}_{H_{t}}^{*}$ is used.
\end{itemize}

\section{Empirical Performance}
\label{section:EmpiricalPerformance}
We assess the empirical performance of the proposed methods through the exercise of predicting the time series returns of foreign exchange rates among major currencies.
Section \ref{section:Data_and_Statistics} describes the data statistics and the outcomes of fitting the QMLE to the DCC-GARCH model.
Section \ref{section:ResidualFitting} discusses the application of a distribution with a correlation adjustment add-in
and correlation matrix decompositions to the residual of the DCC-GARCH model.
We compare these approaches using the out-of-sample likelihood.

\subsection{Data and DCC-GARCH Fitting}
\begin{figure}[H]
  \centering
  \begin{minipage}[t]{0.9\hsize}
    \centering
    \begin{tabular}{cc}
      \begin{minipage}[t]{0.5\hsize}
        \centering
        \includegraphics[keepaspectratio, scale=0.55]{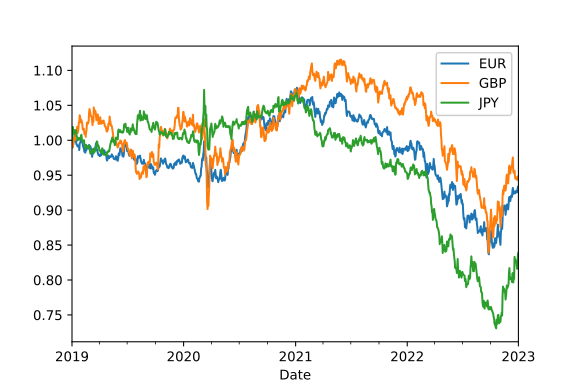}
        \caption{Historical rates for Group1 (EUR, GBP, JPY).}
        \label{figure:C4_MCPriceRatio}
      \end{minipage} &
      \begin{minipage}[t]{0.5\hsize}
        \centering
        \includegraphics[keepaspectratio, scale=0.55]{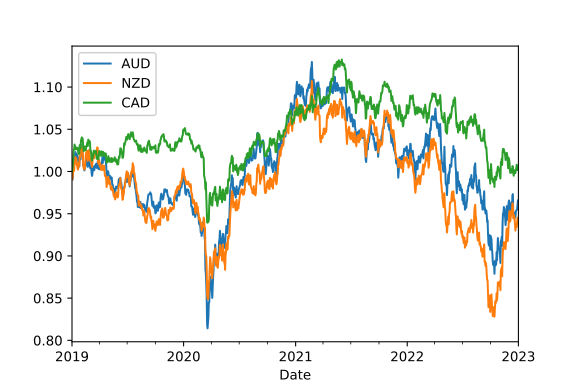}
        \caption{Historical rates for Group2 (AUD, NZD, CAD).}
        \label{figure:C4_RCPriceRatio}
        \end{minipage}\\
      \begin{minipage}[t]{0.5\hsize}
        \centering
        \includegraphics[keepaspectratio, scale=0.55]{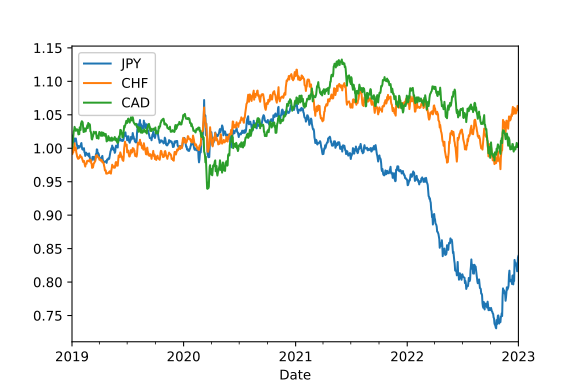}
        \caption{Historical rates for Group3 (AUD, NZD, CAD).}
        \label{figure:C4_NCPriceRatio}
      \end{minipage} &
    \end{tabular}
  \end{minipage}\\
  \begin{minipage}[t]{0.9\hsize}
    {\footnotesize Relative rates of the foreign exchange rates against the USD, with prices from January 1, 2019, as 1.
For JPY, CHF, and CAD, we use the inverse of the commonly quoted exchange rates USD-JPY, USD-CHF, and USD-CAD.
The market disruption caused by the global spread of COVID-19 infection is seen around March 2020. 
Tightening cycles in major advanced economies started in 2022. }
  \end{minipage}
\end{figure}

We use the daily foreign exchange rates of EUR, GBP, JPY, AUD, NZD, CHF, and CAD against the USD from December 31, 2018, to December 29, 2022,
obtained from Bloomberg.\footnote{Supported by the Center for Advanced Research in Finance (CARF).}.
For JPY, CHF, and CAD, we use the inverse of the commonly quoted rates of USD-JPY, USD-CHF, and USD-CAD.
These foreign exchange rates are divided into three groups as follows: Group1 (EUR, GBP, JPY): High liquidity; Group2 (AUD, NZD, CAD): High correlation; Group3 (JPY, CHF, CAD): Low correlation.
The relative rates for January 1, 2019, as 1 are shown in Figures \ref{figure:C4_MCPriceRatio}, \ref{figure:C4_RCPriceRatio}, and \ref{figure:C4_NCPriceRatio}.
For these foreign exchange rates, we use the logarithmic returns from the previous observation date.
We consider January 1, 2019, to December 30, 2021, as the in-sample period (783 data points) and, thereafter, as the out-of-sample period (259 data points).

\label{section:Data_and_Statistics}

The statistics of the logarithmic returns for the in-sample period are shown in Table \ref{table:C4_LogRateIsStatistics},
whereas those for the out-of-sample period are listed in Table \ref{table:C4_LogRateOosStatistics}.
Tables \ref{table:C4_MCIsStatistics2}-\ref{table:C4_NCOosStatistics2} show the linear and rank correlation coefficients of the logarithmic returns for each group.
The values in the upper right of the tables denote linear correlations, whereas those on the lower left denote rank correlations.
The correlations among the foreign exchange rates in Group2 are higher than those in Group1, whereas those in Group3 are lower than those in Group1.
The in-sample period includes a period of market disruption (around March 2020; see \cite{mazur2021covid}) owing to the global spread of COVID-19.
The out-of-sample period corresponds to the period when central banks in major advanced economies tightend their cycles owing to rising global inflation,
which led to increased volatility (\cite{drehmann2023papers}).

\begin{table}[H]
  \centering
  \begin{tabular}{c}
    \begin{minipage}[t]{1.0\hsize}
      \centering
      \caption{Statistics of the logarithmic returns (in-sample period).}
      \label{table:C4_LogRateIsStatistics}
      \begin{tabular}{rrrrrrrr}
        \toprule
        {} &          EUR &          GBP &          JPY & AUD & NZD & CHF & CAD \\
        \midrule
        mean  &      -0.0000 &       0.0001 &      -0.0001 &       0.0000 &       0.0000 &       0.0001 &       0.0001 \\
        std   &       0.0038 &       0.0055 &       0.0043 &       0.0060 &       0.0059 &       0.0040 &       0.0043 \\
        skew  &      -0.2023 &      -0.2342 &      -0.4410 &      -0.6529 &      -0.5653 &      -0.1009 &      -0.3280 \\
        kurt  &       1.8758 &       4.6390 &      10.9482 &       4.9678 &       2.6769 &       1.6353 &       2.6504 \\
        min   &      -0.0206 &      -0.0378 &      -0.0315 &      -0.0390 &      -0.0353 &      -0.0181 &      -0.0210 \\
        25\%   &      -0.0022 &      -0.0031 &      -0.0023 &      -0.0033 &      -0.0033 &      -0.0022 &      -0.0022 \\
        50\%   &       0.0000 &       0.0000 &      -0.0001 &       0.0001 &       0.0002 &      -0.0001 &       0.0002 \\
        75\%   &       0.0024 &       0.0033 &       0.0019 &       0.0035 &       0.0038 &       0.0024 &       0.0025 \\
        max   &       0.0146 &       0.0270 &       0.0292 &       0.0203 &       0.0188 &       0.0158 &       0.0189 \\
        \bottomrule
      \end{tabular}
    \end{minipage}\\\\
    \begin{minipage}[t]{1.0\hsize}
      \centering
      \caption{Statistics of the logarithmic returns (out-of-sample period).}
      \label{table:C4_LogRateOosStatistics}
      \begin{tabular}{rrrrrrrr}
        \toprule
        {} &          EUR &          GBP &          JPY & AUD & NZD & CHF & CAD \\
        \midrule
        mean  &      -0.0002 &      -0.0004 &      -0.0005 &      -0.0002 &      -0.0003 &      -0.0000 &      -0.0003 \\
        std   &       0.0062 &       0.0077 &       0.0075 &       0.0085 &       0.0079 &       0.0058 &       0.0052 \\
        skew  &       0.2766 &      -0.0605 &       1.1219 &       0.1198 &       0.1034 &       1.0888 &       0.2692 \\
        kurt  &       0.4859 &       2.6869 &       5.0777 &       0.6906 &       0.6196 &       3.1189 &       0.6135 \\
        min   &      -0.0153 &      -0.0364 &      -0.0210 &      -0.0237 &      -0.0231 &      -0.0131 &      -0.0138 \\
        25\%   &      -0.0043 &      -0.0043 &      -0.0043 &      -0.0055 &      -0.0050 &      -0.0039 &      -0.0037 \\
        50\%   &      -0.0001 &      -0.0002 &      -0.0009 &      -0.0001 &      -0.0003 &      -0.0006 &      -0.0004 \\
        75\%   &       0.0031 &       0.0034 &       0.0026 &       0.0049 &       0.0046 &       0.0030 &       0.0030 \\
        max   &       0.0211 &       0.0310 &       0.0386 &       0.0288 &       0.0270 &       0.0282 &       0.0196 \\
        \bottomrule
      \end{tabular}
    \end{minipage}
  \end{tabular}\\
  \begin{minipage}[t]{0.9\hsize}
    {\footnotesize Statistics on the logarithmic returns of the foreign exchange rates.
    Each row represents mean: mean, std: standard deviation, skew: skewness, kurt: kurtosis of Fisher's definition (normal is $0$), min: minimum value, 25\%, 50\%, 75\%: lower 25, 50, and 75 percentile points, respectively, max: maximum value. }
  \end{minipage}
\end{table}

\begin{table}[H]
  \centering
  \begin{minipage}[t]{0.9\hsize}
    \centering
    \begin{tabular}{cc}
      \begin{minipage}[t]{0.5\hsize}
        \caption{Correlation of $\bm{r}_{t}$ for in-sample period in Group1}
        \label{table:C4_MCIsStatistics2}
        \centering
        \begin{tabular}{lrrr}
          \toprule
          {} &    EUR &    GBP &    JPY \\
          \midrule
          EUR & 1.0000 & 0.5556 & 0.4853 \\
          GBP & 0.5533 & 1.0000 & 0.3140 \\
          JPY & 0.3995 & 0.2746 & 1.0000 \\
          \bottomrule
          \end{tabular}
      \end{minipage} &
      \begin{minipage}[t]{0.5\hsize}
        \caption{Correlation of $\bm{r}_{t}$ for out-of-sample period in Group1}
        \label{table:C4_MCOosStatistics2}
        \centering
        \begin{tabular}{lrrr}
          \toprule
          {} &    EUR &    GBP &    JPY \\
          \midrule
          EUR & 1.0000 & 0.7853 & 0.4246 \\
          GBP & 0.7784 & 1.0000 & 0.4750 \\
          JPY & 0.4202 & 0.4590 & 1.0000 \\
          \bottomrule
          \end{tabular}
      \end{minipage}\\
      \begin{minipage}[t]{0.5\hsize}
        \caption{Correlation of $\bm{r}_{t}$ for in-sample period in Group2}
        \label{table:C4_RCIsStatistics2}
        \centering
        \begin{tabular}{lrrr}
          \toprule
          {} &    AUD &    NZD &    CAD \\
          \midrule
          AUD & 1.0000 & 0.8719 & 0.6828 \\
          NZD & 0.8467 & 1.0000 & 0.6247 \\
          CAD & 0.6477 & 0.5970 & 1.0000 \\
          \bottomrule
          \end{tabular}     
      \end{minipage} &
      \begin{minipage}[t]{0.5\hsize}
        \caption{Correlation of $\bm{r}_{t}$ for out-of-sample period in Group2}
        \label{table:C4_RCOosStatistics2}
        \centering
        \begin{tabular}{lrrr}
          \toprule
          {} &    AUD &    NZD &    CAD \\
          \midrule
          AUD & 1.0000 & 0.9198 & 0.8166 \\
          NZD & 0.8982 & 1.0000 & 0.7493 \\
          CAD & 0.7872 & 0.6829 & 1.0000 \\
          \bottomrule
          \end{tabular}
      \end{minipage}\\
      \begin{minipage}[t]{0.5\hsize}
        \caption{Correlation of $\bm{r}_{t}$ for in-sample period in Group3}
        \label{table:C4_NCIsStatistics2}
        \centering
        \begin{tabular}{lrrr}
          \toprule
          {} &    JPY &    CHF &     CAD \\
          \midrule
          JPY & 1.0000 & 0.5911 & -0.0298 \\
          CHF & 0.5104 & 1.0000 &  0.2664 \\
          CAD & 0.0418 & 0.3054 &  1.0000 \\
          \bottomrule
          \end{tabular}
      \end{minipage} &
      \begin{minipage}[t]{0.5\hsize}
        \caption{Correlation of $\bm{r}_{t}$ for out-of-sample period in Group3}
        \label{table:C4_NCOosStatistics2}
        \centering
        \begin{tabular}{lrrr}
          \toprule
          {} &    JPY &    CHF &    CAD \\
          \midrule
          JPY & 1.0000 & 0.5325 & 0.3309 \\
          CHF & 0.5212 & 1.0000 & 0.5629 \\
          CAD & 0.2913 & 0.5805 & 1.0000 \\
          \bottomrule
          \end{tabular}
      \end{minipage}
    \end{tabular}
  \end{minipage}\\
  \begin{minipage}[t]{0.9\hsize}
    {\footnotesize Correlation of logarithmic returns of the foreign exchange rates. The values in the upper right of the table denote linear correlations, whereas the values in the lower left denote rank correlations. }
  \end{minipage}
\end{table}

Table \ref{table:C4_GarchFitting} presents the results of the parameter estimates of the GARCH model for the in-sample period.
In parameter estimation, the constraint $\sigma_{i,0}=\bar{\sigma}_{i}$ is set, which assumes that $\mathcal{F}=\mathcal{F}_{0}$.
The statistics of the GARCH residual for the in-sample period are shown in Table \ref{table:C4_GarchFittingXiISStatistics1}, and those for the out-of-sample period are shown in Table \ref{table:C4_GarchFittingXiOOSStatistics1}.
Comparing Tables \ref{table:C4_LogRateIsStatistics} and \ref{table:C4_LogRateOosStatistics},
the kurtosis of the GARCH residual is much lower than that of the logarithmic returns.
However, they are still far from $0$, which motivates us to use a distribution other than a Gaussian distribution.
Tables \ref{table:C4_MCXiIsStatistics2}-\ref{table:C4_NCXiOosStatistics2} show the linear and rank correlations of the GARCH residual within each group.
The values in the upper right of the tables denote linear correlations, whereas those on the lower left denote rank correlations.
By comparing Tables \ref{table:C4_MCIsStatistics2}-\ref{table:C4_NCOosStatistics2},
the correlations between the assets in the logarithmic returns are close to those in the GARCH residual.

\begin{table}[H]
  \centering
  \begin{minipage}[t]{0.9\hsize}
    \caption{Estimated parameters of the GARCH model.}
    \label{table:C4_GarchFitting}
    \centering
    \begin{tabular}{lcccc}
      \toprule
      Currency & $\omega_{i}$ &  $\alpha_{i}$ & $\beta_{i}$ & $\sigma_{0}=\bar{\sigma}$ \\
      \midrule
      EUR &  5.410e-07 & 0.0653 & 0.8970 & 0.0038 \\
      GBP &  3.639e-06 & 0.1327 & 0.7355 & 0.0053 \\
      JPY &  1.858e-06 & 0.1168 & 0.7601 & 0.0039 \\
      AUD &  7.539e-07 & 0.0632 & 0.9161 & 0.0060 \\
      NZD &  8.278e-07 & 0.0433 & 0.9320 & 0.0058 \\
      CHF &  1.245e-06 & 0.0763 & 0.8449 & 0.0040 \\
      CAD &  4.418e-07 & 0.0568 & 0.9187 & 0.0042 \\
      \bottomrule
    \end{tabular}
  \end{minipage}\\
  \begin{minipage}[t]{0.9\hsize}
    {\footnotesize Parameter estimates of the GARCH model for the in-sample period.
    The constraint $\sigma_{i,0}=\bar{\sigma}_{i}$ is set in estimation, which assumes that $\mathcal{F}=\mathcal{F}_{0}$. }
  \end{minipage}
\end{table}

\begin{table}[H]
  \centering
  \begin{tabular}{c}
    \begin{minipage}[t]{1.0\hsize}
      \caption{Statistics of the GARCH residual $\xi_{i,t}$ (in-sample period).}
      \label{table:C4_GarchFittingXiISStatistics1}
      \centering
      \begin{tabular}{llllllll}
        \toprule
        {} &          EUR &          GBP &          JPY &          AUD &          NZD &          CHF &          CAD \\
        \midrule
        mean  &      -0.0018 &       0.0194 &      -0.0125 &      -0.0099 &      -0.0057 &       0.0329 &       0.0144 \\
        std   &       0.9936 &       0.9993 &       0.9980 &       0.9963 &       1.0004 &       0.9978 &       0.9988 \\
        skew  &      -0.0374 &       0.1062 &       0.1332 &      -0.5159 &      -0.5197 &       0.0908 &      -0.4234 \\
        kurt  &       0.5593 &       0.8692 &       2.5923 &       2.5148 &       1.6276 &       0.7898 &       2.1602 \\
        min   &      -3.4477 &      -3.3456 &      -4.6134 &      -6.5228 &      -5.4400 &      -3.4049 &      -6.2745 \\
        25\%   &      -0.6092 &      -0.6205 &      -0.5816 &      -0.5942 &      -0.5984 &      -0.5472 &      -0.5661 \\
        50\%   &       0.0000 &       0.0000 &      -0.0249 &       0.0203 &       0.0320 &      -0.0251 &       0.0421 \\
        75\%   &       0.6326 &       0.6605 &       0.5218 &       0.6392 &       0.6394 &       0.6010 &       0.6017 \\
        max   &       3.5969 &       4.3399 &       4.5702 &       3.0016 &       2.8206 &       3.8826 &       2.7787 \\
        \bottomrule
      \end{tabular}
    \end{minipage}\\\\
    \begin{minipage}[t]{1.0\hsize}
      \caption{Statistics of the GARCH residual $\xi_{i,t}$ (out-of-sample period).}
      \label{table:C4_GarchFittingXiOOSStatistics1}
      \centering
      \begin{tabular}{llllllll}
        \toprule
        {} &          EUR &          GBP &          JPY &          AUD &          NZD &          CHF &          CAD \\
        \midrule
        mean  &      -0.0701 &      -0.1097 &      -0.1278 &      -0.0403 &      -0.0565 &      -0.0380 &      -0.0622 \\
        std   &       1.1746 &       1.1815 &       1.3875 &       1.1135 &       1.1186 &       1.2025 &       1.0757 \\
        skew  &       0.0175 &      -0.6753 &       0.9430 &      -0.0236 &      -0.0242 &       0.6097 &       0.0866 \\
        kurt  &       0.1239 &       3.0692 &       4.1885 &       0.2851 &       0.0940 &       1.4806 &       0.4679 \\
        min   &      -3.1635 &      -6.3962 &      -3.7266 &      -3.2347 &      -3.5168 &      -3.0600 &      -3.3594 \\
        25\%   &      -0.8215 &      -0.7701 &      -0.8826 &      -0.7138 &      -0.8188 &      -0.8267 &      -0.7574 \\
        50\%   &      -0.0278 &      -0.0410 &      -0.1793 &      -0.0205 &      -0.0412 &      -0.1311 &      -0.0917 \\
        75\%   &       0.6503 &       0.5268 &       0.5690 &       0.6859 &       0.6966 &       0.6843 &       0.6817 \\
        max   &       3.3738 &       3.0466 &       7.4362 &       3.4059 &       3.4060 &       5.3511 &       3.8502 \\
        \bottomrule
      \end{tabular}
    \end{minipage}
  \end{tabular}\\
  \begin{minipage}[t]{0.9\hsize}
    {\footnotesize Statistics on the GARCH residual of foreign exchange rates.
    Each row represents mean: mean, std: standard deviation, skew: skewness, kurt: kurtosis of Fisher's definition (normal is $0$), min: minimum value, 25\%, 50\%, 75\%: lower 25, 50, and 75 percentile points, respectively, max: maximum value. }
  \end{minipage}
\end{table}

\begin{table}[H]
  \centering
  \begin{minipage}[t]{0.9\hsize}
    \centering
    \begin{tabular}{cc}
      \begin{minipage}[t]{0.5\hsize}
        \caption{Correlation of GARCH residual $\xi_{i}$ for in-sample period in Group1}
        \label{table:C4_MCXiIsStatistics2}
        \centering
        \begin{tabular}{lrrr}
          \toprule
          {} &    EUR &    GBP &    JPY \\
          \midrule
          EUR & 1.0000 & 0.5410 & 0.4116 \\
          GBP & 0.5606 & 1.0000 & 0.2628 \\
          JPY & 0.3954 & 0.2752 & 1.0000 \\
          \bottomrule
          \end{tabular}
      \end{minipage} &
      \begin{minipage}[t]{0.5\hsize}
        \caption{Correlation of GARCH residual $\xi_{i}$ for out-of-sample period in Group1}
        \label{table:C4_MCXiOosStatistics2}
        \centering
        \begin{tabular}{lrrr}
          \toprule
          {} &    EUR &    GBP &    JPY \\
          \midrule
          EUR & 1.0000 & 0.7567 & 0.3718 \\
          GBP & 0.7760 & 1.0000 & 0.4237 \\
          JPY & 0.3987 & 0.4403 & 1.0000 \\
          \bottomrule
          \end{tabular}
      \end{minipage}\\
      \begin{minipage}[t]{0.5\hsize}
        \caption{Correlation of GARCH residual $\xi_{i}$ for in-sample period in Group2}
        \label{table:C4_RCXiIsStatistics2}
        \centering
        \begin{tabular}{lrrr}
          \toprule
          {} &    AUD &    NZD &    CAD \\
          \midrule
          AUD & 1.0000 & 0.8577 & 0.6443 \\
          NZD & 0.8431 & 1.0000 & 0.5861 \\
          CAD & 0.6382 & 0.5926 & 1.0000 \\
          \bottomrule
          \end{tabular}
      \end{minipage} &
      \begin{minipage}[t]{0.5\hsize}
        \caption{Correlation of GARCH residual $\xi_{i}$ for out-of-sample period in Group2}
        \label{table:C4_RCXiOosStatistics2}
        \centering
        \begin{tabular}{lrrr}
          \toprule
          {} &    AUD &    NZD &    CAD \\
          \midrule
          AUD & 1.0000 & 0.9178 & 0.8069 \\
          NZD & 0.9042 & 1.0000 & 0.7329 \\
          CAD & 0.7856 & 0.6879 & 1.0000 \\
          \bottomrule
          \end{tabular}
      \end{minipage}\\
      \begin{minipage}[t]{0.5\hsize}
        \caption{Correlation of GARCH residual $\xi_{i}$ for in-sample period in Group3}
        \label{table:C4_NCXiIsStatistics2}
        \centering
        \begin{tabular}{lrrr}
          \toprule
          {} &    JPY &    CHF &     CAD \\
          \midrule
          JPY & 1.0000 & 0.5435 & -0.0103 \\
          CHF & 0.5050 & 1.0000 &  0.2593 \\
          CAD & 0.0461 & 0.3140 &  1.0000 \\
          \bottomrule
          \end{tabular}
      \end{minipage} &
      \begin{minipage}[t]{0.5\hsize}
        \caption{Correlation of GARCH residual $\xi_{i}$ for out-of-sample period in Group3}
        \label{table:C4_NCXiOosStatistics2}
        \centering
        \begin{tabular}{lrrr}
          \toprule
          {} &    JPY &    CHF &    CAD \\
          \midrule
          JPY & 1.0000 & 0.5124 & 0.3050 \\
          CHF & 0.5301 & 1.0000 & 0.5505 \\
          CAD & 0.2837 & 0.5731 & 1.0000 \\
          \bottomrule
          \end{tabular}
      \end{minipage}
    \end{tabular}
  \end{minipage}\\
  \begin{minipage}[t]{0.9\hsize}
    {\footnotesize Correlations of the GARCH residual for each foreign exchange rate. The values in the upper right of the table denote linear correlations, whereas the values in the lower left denote rank correlations. }
  \end{minipage}
\end{table}

Table \ref{table:C4_CcyDCCResult} presents the parameter estimates of the DCC model using the second-stage formula \eqref{formula:DCCCalibration} for each group.
In the estimation, we restrict $Q_{0}=\bar{Q}$; that is, the initial value of the model's variance-covariance matrix coincides with the unconditional correlation matrix.
$\mathcal{F}=\mathcal{F}_{0}$ is assumed as in the GARCH model.

\begin{figure}[H]
  \centering
  \begin{minipage}[t]{0.9\hsize}
    \centering
    \begin{tabular}{cc}
      \begin{minipage}[t]{0.5\hsize}
        \centering
        \includegraphics[keepaspectratio, scale=0.55]{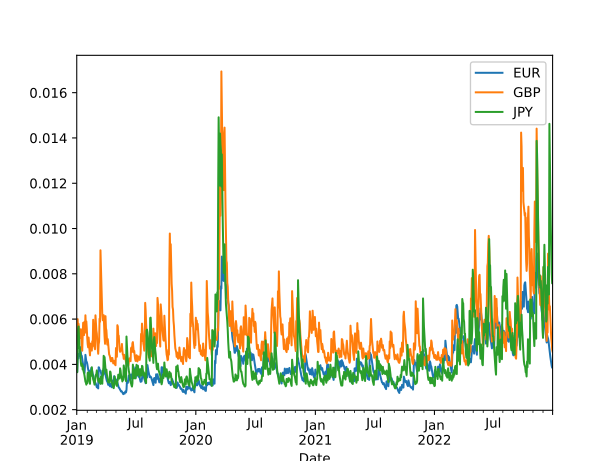}
        \caption{Group1 EUR, GBP, JPY. }
        \label{figure:C4_MCGarchSigma}
      \end{minipage} &
      \begin{minipage}[t]{0.5\hsize}
        \centering
        \includegraphics[keepaspectratio, scale=0.55]{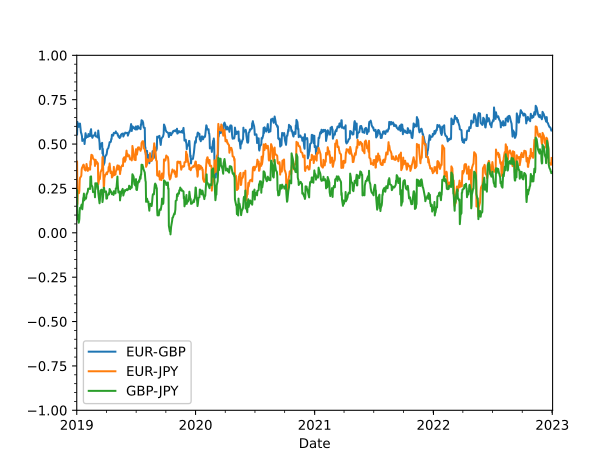}
        \caption{Group1 EUR, GBP, JPY. }
        \label{figure:C4_MCDCCRho}
      \end{minipage}\\
      \begin{minipage}[t]{0.5\hsize}
        \centering
        \includegraphics[keepaspectratio, scale=0.55]{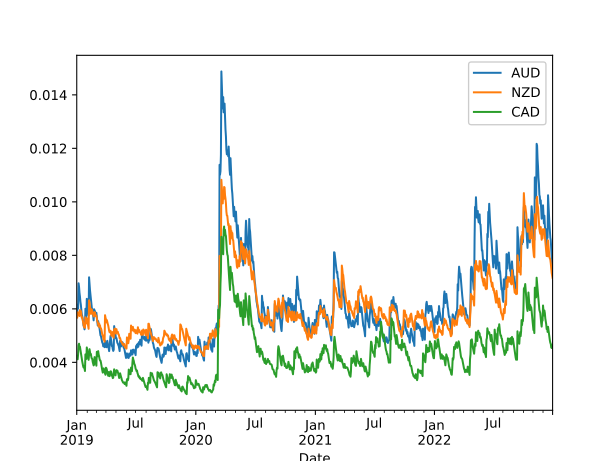}
        \caption{Group2 AUD, NZD, CAD. }
        \label{figure:C4_RCGarchSigma}
      \end{minipage} &
      \begin{minipage}[t]{0.5\hsize}
        \centering
        \includegraphics[keepaspectratio, scale=0.55]{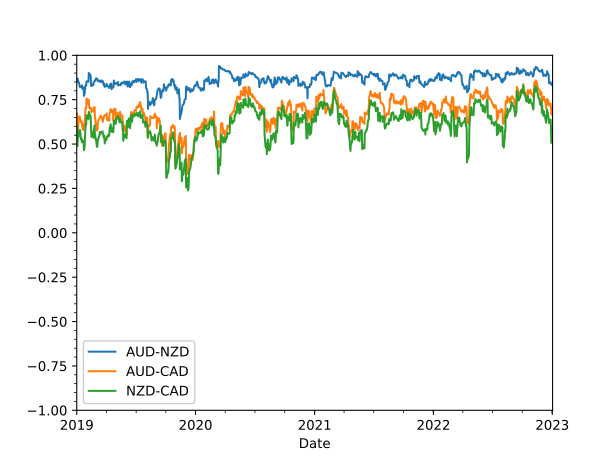}
        \caption{Group2 AUD, NZD, CAD. }
        \label{figure:C4_RCDCCRho}
      \end{minipage}\\
      \begin{minipage}[t]{0.5\hsize}
        \centering
        \includegraphics[keepaspectratio, scale=0.55]{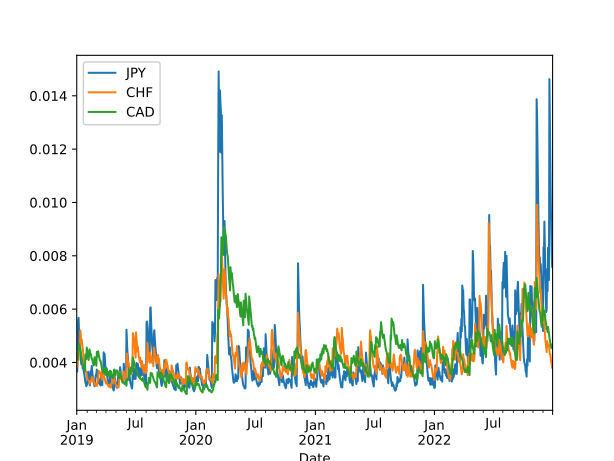}
        \caption{Group3 JPY, CHF, CAD. }
        \label{figure:C4_NCGarchSigma}
      \end{minipage} &
      \begin{minipage}[t]{0.5\hsize}
        \centering
        \includegraphics[keepaspectratio, scale=0.55]{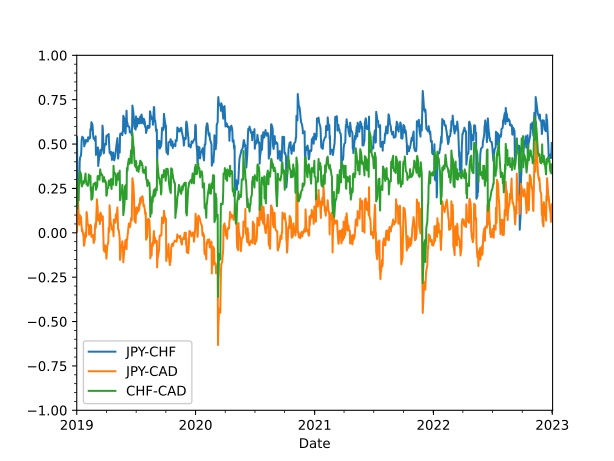}
        \caption{Group3 JPY, CHF, CAD. }
        \label{figure:C4_NCDCCRho}
      \end{minipage}
    \end{tabular}
  \end{minipage}\\
  \begin{minipage}[t]{0.9\hsize}
    {\footnotesize Figures \ref{figure:C4_MCGarchSigma}, \ref{figure:C4_RCGarchSigma}, and \ref{figure:C4_NCGarchSigma} are the time series of volatility $\sigma_{i,t}$ estimated by the GARCH model,
    and Figures \ref{figure:C4_MCDCCRho}, \ref{figure:C4_RCDCCRho}, and \ref{figure:C4_NCDCCRho} are the time series of correlation $R_{t}$ estimated by the DCC model. }
  \end{minipage}
\end{figure}

Figures \ref{figure:C4_MCGarchSigma}, \ref{figure:C4_RCGarchSigma}, and \ref{figure:C4_NCGarchSigma} show the time series of volatility $\sigma_{i,t}$ estimated by the GARCH model.
Figures \ref{figure:C4_MCDCCRho}, \ref{figure:C4_RCDCCRho}, and \ref{figure:C4_NCDCCRho} show the time series of correlation $R_{t}$ estimated by the DCC model.
For each group, we observe periods of high volatility in the first half of 2020 and the second half of 2022.
No significant fluctuations linked to the volatility levels are observed for these periods in the first half of 2020 or the second half of 2022 with respect to correlations.
However, the correlations vary throughout the period, and small spikes are observed.
Not excluding the dynamic correlation model is appropriate when treating exchange rate returns.

\begin{table}[H]
  \centering
  \begin{minipage}[t]{0.9\hsize}
  \centering
    \caption{Estimated parameters of DCC model in each group}
    \label{table:C4_CcyDCCResult}
    \begin{tabular}{cccc}
      \toprule
      Asset Group & $a$ & $b$ & $Q_{0}=\bar{Q}$\\
      \midrule
      Group1 & 0.03103 & 0.8815 & {\small$\begin{pmatrix}1 & 0.5532 & 0.4017\\ 0.5532 & 1 & 0.2397\\ 0.4017 & 0.2397 & 1\end{pmatrix}$} \\
      Group2 & 0.04567 & 0.9037 & {\small$\begin{pmatrix}1 & 0.8521 & 0.6752\\ 0.8521 & 1 & 0.6063\\ 0.6752 & 0.6063 & 1\end{pmatrix}$} \\
      Group3 & 0.05784 & 0.7979 & {\small$\begin{pmatrix}1 & 0.5417 & 0.0048\\ 0.5417 & 1 & 0.2964\\ 0.0048 & 0.2964 & 1\end{pmatrix}$} \\
      \bottomrule
    \end{tabular}
  \end{minipage}
  \begin{minipage}[t]{0.9\hsize}
    {\footnotesize The results of the second stage of DCC estimation by formula \eqref{formula:DCCCalibration} for the GARCH residual $\xi_{t}$ of each group.
    The constraint $Q_{0}=\bar{Q}$ is set in estimation, which assumes $\mathcal{F}=\mathcal{F}_{0}$. }
  \end{minipage}
\end{table}

\subsection{DCC-GARCH Residual Fitting}
\label{section:ResidualFitting}
We compare the methods proposed in this study using DCC-GARCH parameters and residuals.
We examine the statistics of the DCC residuals obtained by the decomposition of the correlation matrix to verify whether filtering out the correlations works.
From the GARCH residual $\bm{\xi}_{t}$ and the correlation matrix $R_{t}$ estimated by the DCC-GARCH model in the previous section,
we calculate the DCC residual $\epsilon_{t}$ using each decomposition method from Section \ref{section:correlation_matrix_decomposition}
and obtain the historical correlation of the DCC residual for the in-sample period.
We also calculated the two-sided $95\%$ confidence interval for the historical correlation using the bootstrap method with 10,000 resamplings.
In addition, the confidence intervals of GARCH residual $\bm{\xi}_{t}$, whose correlations are listed in Tables \ref{table:C4_MCXiIsStatistics2}, \ref{table:C4_RCXiIsStatistics2}, and \ref{table:C4_NCXiIsStatistics2},
are calculated as ``NoDCC.'' We examine NoDCC here because several distributions fit the copula-GARCH model and compared them in a later validation.
The results of the historical correlations and confidence intervals are presented in Tables \ref{table:C4_Group1CorrConfidentialInterval}-\ref{table:C4_Group3CorrConfidentialInterval}.

\begin{table}[H]
  \centering
  \begin{minipage}[t]{1.0\hsize}
    \caption{Correlations and confidence intervals for the residual in Group1}
    \label{table:C4_Group1CorrConfidentialInterval}
    \centering
    \begin{tabular}{lrrrrrr}
      \hline
      {} &  NoDCC &    Sqrt &   Sqrt2 &  Cholesky &  Eigen &  Eigen2 \\
      \hline
      $\rho_{12}$       & 0.5410 & -0.0144 & -0.0128 &    0.0013 &  0.0043 &  -0.0255 \\
      $\rho_{12}$ Lower & 0.4896 & -0.0848 & -0.0841 &   -0.0687 & -0.0875 &  -0.1150 \\
      $\rho_{12}$ Upper & 0.5897 &  0.0568 &  0.0594 &    0.0709 &  0.0948 &   0.0633 \\
      \hline
      $\rho_{13}$       & 0.4116 &  0.0075 &  0.0121 &    0.0097 & -0.0122 &  -0.0060 \\
      $\rho_{13}$ Lower & 0.3412 & -0.0754 & -0.0706 &   -0.0775 & -0.0923 &  -0.0876 \\
      $\rho_{13}$ Upper & 0.4785 &  0.0901 &  0.0944 &    0.0961 &  0.0688 &   0.0755 \\
      \hline
      $\rho_{23}$       & 0.2628 &  0.0114 &  0.0136 &    0.0111 &  0.0132 &   0.0377 \\
      $\rho_{23}$ Lower & 0.1905 & -0.0666 & -0.0638 &   -0.0670 & -0.0721 &  -0.0478 \\
      $\rho_{23}$ Upper & 0.3337 &  0.0903 &  0.0923 &    0.0909 &  0.0981 &   0.1229 \\
      \hline
      \end{tabular}
  \end{minipage}\\
  \begin{minipage}[t]{1.0\hsize}
    \caption{Correlations and confidence intervals for the residual in Group2}
    \label{table:C4_Group2CorrConfidentialInterval}
    \centering
    \begin{tabular}{lrrrrrr}
      \hline
      {} &  NoDCC &    Sqrt &   Sqrt2 &  Cholesky &  Eigen &  Eigen2 \\
      \hline
      $\rho_{12}$       & 0.8577 &  0.0081 &  0.0164 &    0.0259 & -0.0035 &  -0.0148 \\
      $\rho_{12}$ Lower & 0.8327 & -0.0747 & -0.0666 &   -0.0487 & -0.0955 &  -0.1028 \\
      $\rho_{12}$ Upper & 0.8798 &  0.0928 &  0.1001 &    0.1018 &  0.0891 &   0.0721 \\
      \hline
      $\rho_{13}$       & 0.6443 & -0.0147 & -0.0039 &   -0.0006 & -0.0193 &  -0.0256 \\
      $\rho_{13}$ Lower & 0.6005 & -0.0856 & -0.0761 &   -0.0814 & -0.1065 &  -0.1124 \\
      $\rho_{13}$ Upper & 0.6872 &  0.0551 &  0.0671 &    0.0804 &  0.0672 &   0.0603 \\
      \hline
      $\rho_{23}$       & 0.5861 &  0.0046 &  0.0068 &   -0.0087 & -0.0089 &  -0.0287 \\
      $\rho_{23}$ Lower & 0.5335 & -0.0684 & -0.0651 &   -0.0812 & -0.0870 &  -0.1056 \\
      $\rho_{23}$ Upper & 0.6367 &  0.0804 &  0.0815 &    0.0668 &  0.0676 &   0.0472 \\
      \hline
      \end{tabular}
  \end{minipage}\\
  \begin{minipage}[t]{1.0\hsize}
    \caption{Correlations and confidence intervals for the residual in Group3}
    \label{table:C4_Group3CorrConfidentialInterval}
    \centering
    \begin{tabular}{lrrrrrr}
      \hline
      {} &   NoDCC &    Sqrt &   Sqrt2 &  Cholesky &  Eigen &  Eigen2 \\
      \hline
      $\rho_{12}$       &  0.5435 &  0.0107 &  0.0131 &    0.0097 &  0.0482 &   0.0570 \\
      $\rho_{12}$ Lower &  0.4790 & -0.0794 & -0.0774 &   -0.0712 & -0.0550 &  -0.0388 \\
      $\rho_{12}$ Upper &  0.6026 &  0.1006 &  0.1034 &    0.0904 &  0.1528 &   0.1528 \\\hline
      $\rho_{13}$       & -0.0103 & -0.0086 & -0.0117 &   -0.0206 & -0.0078 &   0.0369 \\
      $\rho_{13}$ Lower & -0.1104 & -0.0985 & -0.1046 &   -0.1253 & -0.0883 &  -0.0525 \\
      $\rho_{13}$ Upper &  0.0865 &  0.0814 &  0.0804 &    0.0810 &  0.0737 &   0.1272 \\\hline
      $\rho_{23}$       &  0.2593 & -0.0114 & -0.0051 &    0.0016 &  0.0078 &   0.0254 \\
      $\rho_{23}$ Lower &  0.1630 & -0.0966 & -0.0882 &   -0.0741 & -0.0742 &  -0.0664 \\
      $\rho_{23}$ Upper &  0.3487 &  0.0738 &  0.0780 &    0.0769 &  0.0908 &   0.1169 \\
      \hline
      \end{tabular}
  \end{minipage}\\
  \begin{minipage}[t]{0.9\hsize}
    {\footnotesize 
    Historical correlations among the elements of the GARCH or DCC-GARCH residual in the in-sample period and $95\%$ confidence intervals using the bootstrap method.
    NoDCC columns are the results with GARCH residual $\bm{\xi}_{t}$; Sqrt, Sqrt2, Cholesky, Eigen, and Eigen2 are the results with DCC residual $\bm{\epsilon}_{t}$ using the decomposition in Section \ref{section:correlation_matrix_decomposition};
    $\rho_{i,j}$ rows represent the historical correlations between the $i$-th and $j$-th elements; $\rho_{ij}$ Lower and $\rho_{ij}$ Upper are the lower and upper $2.5\%$ tile points, respectively. }
  \end{minipage}
\end{table}

Columns Sqrt, Sqrt2, Cholesky, Eigen, and Eigen2 correspond to the decomposition methods described in Section \ref{section:correlation_matrix_decomposition}.
NoDCC columns correspond to the historical correlations and their confidence intervals for GARCH residual $\xi_{t}$.
The correlations of NoDCC are significantly non-zero, except for $\rho_{13}$ in Group3,
whereas the correlation is not rejected as zero for DCC residuals when any decomposition method is used.
From these results, the GARCH residual of the data has a non-zero correlation, and each method is consistent with the model hypothesis that the DCC residual $\bm{\xi}_{t}$ is linearly uncorrelated.

To observe higher-order dependency among the residuals, we compute the 2-2 order cokurtosis using
\begin{eqnarray}
  K(X,Y)=\frac{\mathbb{E}\left[(X-\mathbb{E}[X])^{2}(Y-\mathbb{E}[Y])^{2}\right]}{\mathbb{E}\left[(X-\mathbb{E}[X])^{2}\right]\mathbb{E}\left[(Y-\mathbb{E}[Y])^{2}\right]}
\end{eqnarray}
and list them in Table \ref{table:C4_Cokurtosis}.
Group$j$-$xy$ denotes the cokurtosis of the $x$-th and $y$-th elements of the DCC residual $\bm{\epsilon}_{t}$ (columns Sqrt, Sqrt2, Cholesky, Eigen, and Eigen2) or GARCH residual $\bm{\xi}_{t}$ (column NoDCC) in Group $j$.
This value is $1$ under the assumption of independence; however, some values deviate from $1$.
In such cases, selecting distributions with uncorrelated and higher-order dependencies are appropriate.
Furthermore, the 2-2 cokurtosis takes different values depending on the decomposition method.
This means that the distribution of DCC residuals differs depending on the decomposition.
Therefore, we must select a decomposition method that properly models the DCC residual.
\begin{table}[H]
  \centering
  \begin{minipage}[t]{1.0\hsize}
    \caption{Cokurtosis of GARCH and DCC Residuals}
    \label{table:C4_Cokurtosis}
    \centering
    \begin{tabular}{lrrrrrr}
      \hline
      {} &  NoDCC &   Sqrt &  Sqrt2 &  Cholesky &  Eigen &  Eigen2 \\
      \hline
      Group1-12 & 1.5470 & 1.0330 & 1.0648 &    0.9971 &  1.7051 &   1.6320 \\
      Group1-13 & 1.8226 & 1.3951 & 1.3895 &    1.5472 &  1.3334 &   1.3669 \\
      Group1-23 & 1.3156 & 1.2509 & 1.2361 &    1.2760 &  1.4904 &   1.4711 \\\hline
      Group2-12 & 4.2611 & 1.4477 & 1.4381 &    1.1688 &  1.7577 &   1.5769 \\
      Group2-13 & 2.1806 & 0.9982 & 1.0363 &    1.3291 &  1.5694 &   1.5431 \\
      Group2-23 & 1.8723 & 1.1310 & 1.0920 &    1.1269 &  1.2165 &   1.1805 \\\hline
      Group3-12 & 2.5755 & 1.6759 & 1.6824 &    1.3399 &  2.3698 &   2.0023 \\
      Group3-13 & 2.0579 & 1.6743 & 1.7782 &    2.2805 &  1.3312 &   1.6962 \\
      Group3-23 & 1.6421 & 1.4951 & 1.4167 &    1.1616 &  1.3986 &   1.7551 \\
      \hline
    \end{tabular}
  \end{minipage}\\
  \begin{minipage}[t]{0.9\hsize}
    {\footnotesize 2-2 order cokurtosis for the GARCH and DCC residuals.
    Row Group$j$-$xy$ denotes the cokurtosis of the $x$-th element and the $y$-th element in Group$j$.}
  \end{minipage}
\end{table}

We then compare the distributions estimated using the GARCH and DCC residuals from the previous validation.
The GARCH residual $\bm{\xi_{t}}$ and DCC residual $\bm{\epsilon_{t}}$ are fitted to the distributions described below using the maximum likelihood estimation.
For marginals, we use the skew-$t$ distribution (see \cite{wurtz2006parameter}) with a mean of 0 and a variance of 1.
Since the variance of the skew-$t$ distribution does not exist when the degrees of freedom are equal to or less than 2,
we set the constraint such that the parameter of degrees of freedom $\nu$ is larger than 2.001.
The distributions are composed of a combination of copulas and marginal distributions with and without a correlation adjustment add-in.
\begin{itemize}
  \item (IC): Skew-t marginals and an independent copula\\
The joint distribution $G_{IC}(\bm{x})$ which is synthesized by an independent copula and skew-$t$ distributions.

  \item (CIC): Skew-t marginals and an independent copula with the correlation adjustment add-in\\
The joint distribution $G_{CIC}(\bm{x})$ with the correlation adjustment add-in applied to (IC).
Let $X_{IC}$ be the random variable vector following $G_{IC}(\bm{x})$;
$G_{CIC}(\bm{x})$ is the distribution of random variable vector $X_{CIC}:=L_{J,S_{X_{IC}}}X_{IC}$.

  \item (GC): Skew-t marginals and a Gaussian copula\\
The joint distribution $G_{GC}(\bm{x})$ which is synthesized by a Gaussian copula with the correlation matrix $\Sigma_{G}$ and skew-$t$ distributions.

  \item (CGC): Skew-t marginals and a Gaussian copula with the correlation adjustment add-in\\
The joint distribution $G_{CGC}(\bm{x})$ with the correlation adjustment add-in applied to (GC).
Let $X_{GC}$ be the random variable vector following $G_{GC}(\bm{x})$;
$G_{CGC}(\bm{x})$ is the distribution of random variable vector $X_{CGC}:=L_{J,S_{X_{GC}}}X_{GC}$.

  \item (TC): Skew-t marginals and a $t$ copula\\
The joint distribution $G_{TC}(\bm{x})$ which is synthesized by a $t$ copula with correlation matrix $\Sigma_{G}$, the degrees of freedom $\nu$, and skew-$t$ distributions.

  \item (CTC): Skew-t marginals and a $t$ copula with the correlation adjustment add-in\\
The joint distribution $G_{CTC}(\bm{x})$ with the correlation adjustment add-in applied to (TC).
Let $X_{TC}$ be the random variable vector following $G_{TC}(\bm{x})$;
$G_{CTC}(\bm{x})$ is the distribution of random variable vector $X_{CTC}:=L_{J,S_{X_{TC}}}X_{TC}$.

  \item (PC): Skew-t marginals and a pair copula\\
The joint distribution $G_{PC}(\bm{x})$ which is synthesized by the pair copula composed by following two-dimensional copulas and skew-$t$ distributions.
The elements of the pair copula are
(1) Gaussian copula, (2) Frank copula, (3) Plackett copula, (4) Clayton copula, (5) Clayton copula rotated $90^{\circ}$,
(6) Clayton copula rotated $180^{\circ}$, (7) Clayton copula rotated $270^{\circ}$, (8) Gumbel copula,
(9) Gumbel copula rotated $90^{\circ}$, (10) Gumbel copula rotated $180^{\circ}$, (11) Gumbel copula rotated $270^{\circ}$, and (12) $t$ copula.
Allowing for duplication, if we assume 3 dimensions, there are $5,\!184=12^{3}\times3$ pair copulas composed of these 12 kinds of copulas.

  \item (CPC): Skew-t marginals and a pair copula with the correlation adjustment add-in\\
The joint distribution $G_{CPC}(\bm{x})$ with the correlation adjustment add-in applied to (PC).
Let $X_{PC}$ be the random variable vector following $G_{PC}(\bm{x})$;
$G_{CPC}(\bm{x})$ is the distribution of random variable vector $X_{CPC}:=L_{J,S_{X_{PC}}}X_{PC}$.
As in (PC), we consider 5,184 combinations.

\end{itemize}
The distributions of the residuals NoDCC, Sqrt, Sqrt2, Cholesky, Eigen, and Eigen2 are modeled as 
(IC), (CIC), (GC), (CGC), (TC), (CTC), (PC), and (CPC) and compared.
The copula-GARCH model is applied as a residual distribution to NoDCC (GARCH residual).
When the distribution is applied to DCC residuals, it is called the copula-DCC-GARCH model.

Here, we confirm that the correlation adjustment add-in accurately captures the linear correlations of the underlying distributions.
The linear correlations of the estimated distributions are computed using numerical integration to verify
whether they are included in the confidence interval range using the bootstrap method in Tables \ref{table:C4_Group1CorrConfidentialInterval}-\ref{table:C4_Group3CorrConfidentialInterval}.
In Section \ref{section:CopulaProperties}, we construct the density function of the joint distribution
using the copula $C(\bm{u})$ and the marginal distributions $F_{1}(x_{1}),\cdots,F_{N}(x_{N})$ defined as 
$f(x_{1},\cdots,x_{N})=\partial_{\bm{u}}C(F_{1}(x_{1}),\cdots,F_{N}(x_{N}))\prod_{i=1}^{N}f_{i}(x_{i})$.
For a random variable vector $X$ that follows this joint distribution and the appropriate function $A(\bm{x})$,
the expected value $\mathbb{E}\left[A(\bm{X})\right]$ can be transformed into the following formula:
\begin{eqnarray}
  \mathbb{E}\left[A(\bm{X})\right]&=&\int_{\mathbb{R}^{N}}A(\bm{x})c(F_{1}(x_{1}),\cdots,F_{N}(x_{N}))\prod_{i=1}^{N}f_{i}(x_{i})dx_{1}\cdots dx_{N}\nonumber\\
  &=&\int_{\mathbb{R}^{N}}A(F_{1}^{-1}(\Phi(y_{1})),\cdots,F_{N}^{-1}(\Phi(y_{N})))\partial_{\bm{u}}C(\Psi(y_{1}),\cdots,\Psi(y_{N}))\prod_{i=1}^{N}\psi(y_{i})dy_{1}\cdots dy_{N}\\
  &\simeq&\sum_{(y_{1,j},\cdots,y_{N,j})\in D_{G}}A(F_{1}^{-1}(\Psi(y_{1,j})),\cdots,F_{N}^{-1}(\Psi(y_{N,j})))c(\psi(y_{1,j}),\cdots,\psi(y_{N,j}))\prod_{i=1}^{N}\psi(y_{i,j})\Delta_{D_{G}}\bm{y}\label{formula:numerical_intg}
\end{eqnarray}
where $\Psi(x)$ and $\psi(x)$ are the standard normal distribution and density functions.
$D_{G}$ is a grid dividing the area $[-8,8]^{N}$ into 100 sections in each direction,
and $\Delta\bm{y}:=0.16^{N}$ is the volume of one hypercube in grid $D_{G}$.
By using formula \eqref{formula:numerical_intg} and choosing the appropriate $A(\bm{x})$,
we calculate the covariance matrix of the estimated distribution and obtain linear correlations.

Tables \ref{table:C4_CorrelationTestPara} and \ref{table:C4_CorrelationTestPC} show whether the calculated linear correlations are within the confidence interval ranges reported in Tables \ref{table:C4_Group1CorrConfidentialInterval}-\ref{table:C4_Group3CorrConfidentialInterval}.
The vertical axis lists the asset groups and copula types, and the horizontal axis lists the decomposition methods.
``T'' is the case where all three linear correlations are included in the confidence intervals; otherwise, the values are ``F''.
For the pair copula, the number of ``T'' out of 5,184 combinations is given.
Among the parametric copulas examined, the correlation test is accepted except for (IC) of NoDCC.
For the pair copula, the correlation adjustment add-in increases the number of ``T'' in NoDCC.
In almost all other cases, the linear correlations of the estimated distributions are within the confidence intervals.
These results confirm the effectiveness of the correlation adjustment add-in in matching linear correlations, especially for estimating the distribution using highly correlated data.
This result remains unchanged in the out-of-sample analysis, as confirmed later.
For DCC residual data, a consistent distribution is estimated with respect to correlation, even when the correlation adjustment add-in is not used.

\begin{table}[H]
  \centering
  \begin{minipage}[t]{0.9\hsize}
  \centering
  \caption{Comparison of the correlations of estimated distributions and the historical correlations (parametric copulas).}
  \label{table:C4_CorrelationTestPara}
  \begin{tabular}{l|cccccc}
    \hline
    {} & NoDCC & Sqrt & Sqrt2 & Cholesky & Eigen & Eigen2\\
    \hline
Group1-IC & F & T & T & T & T & T\\
Group1-CIC & T & T & T & T & T & T \\
Group1-GC & T & T & T & T & T & T\\
Group1-CGC & T & T & T & T & T & T\\
Group1-TC & T & T & T & T & T & T\\
Group1-CTC & T & T & T & T & T & T\\
\hline
Group2-IC & F & T & T & T & T & T\\
Group2-CIC & T & T & T & T & T & T \\
Group2-GC & T & T & T & T & T & T\\
Group2-CGC & T & T & T & T & T & T\\
Group2-TC & T & T & T & T & T & T\\
Group2-CTC & T & T & T & T & T & T\\
\hline
Group3-IC & F & T & T & T & T & T\\
Group3-CIC & T & T & T & T & T & T \\
Group3-GC & T & T & T & T & T & T\\
Group3-CGC & T & T & T & T & T & T\\
Group3-TC & T & T & T & T & T & T\\
Group3-CTC & T & T & T & T & T & T\\
    \hline
    \end{tabular}
\end{minipage}\\
\begin{minipage}[t]{1.0\hsize}
  {\footnotesize 
  ``T'' is the situation where all linear correlations of the distributions estimated from the GARCH and DCC residuals are included in the confidence intervals of Tables \ref{table:C4_Group1CorrConfidentialInterval}-\ref{table:C4_Group3CorrConfidentialInterval}, and ``F'' is the situation where one or more of them are not included.
  }
\end{minipage}
\end{table}
\begin{table}[H]
  \centering
\begin{minipage}[t]{0.9\hsize}
  \centering
  \caption{Comparison of the correlations of estimated distributions and the historical correlations (pair copulas).}
  \label{table:C4_CorrelationTestPC}
  \begin{tabular}{l|cccccc}
    \hline
    {} & NoDCC & Sqrt & Sqrt2 & Cholesky & Eigen & Eigen2\\
    \hline
Group1-PC & 1164 & 5184 & 5184 & 5184 & 5184 & 5184\\
Group1-CPC & 5179 & 5184 & 5184 & 5184 & 5184 & 5184\\
\hline
Group2-PC & 365 & 5184 & 5184 & 5184 & 5184 & 5184\\
Group2-CPC & 5184 & 5184 & 5184 & 5184 & 5184 & 5184\\
\hline
Group3-PC & 1076 & 5184 & 5184 & 5184 & 5184 & 5183\\
Group3-CPC & 5184 & 5184 & 5184 & 5184 & 5184 & 5184\\
    \hline
    \end{tabular}
\end{minipage}\\
\begin{minipage}[t]{1.0\hsize}
  {\footnotesize 
  The number of all linear correlations of the distributions estimated from the GARCH and DCC residuals included in the confidence intervals of Tables \ref{table:C4_Group1CorrConfidentialInterval}-\ref{table:C4_Group3CorrConfidentialInterval}
  out of 5,184 combinations is given.
  }
\end{minipage}
\end{table}

\begin{figure}[H]
  \centering
  \begin{minipage}[t]{0.9\hsize}
    \centering
    \begin{tabular}{cc}
      \begin{minipage}[t]{0.5\hsize}
        \centering
        \includegraphics[keepaspectratio, scale=0.55]{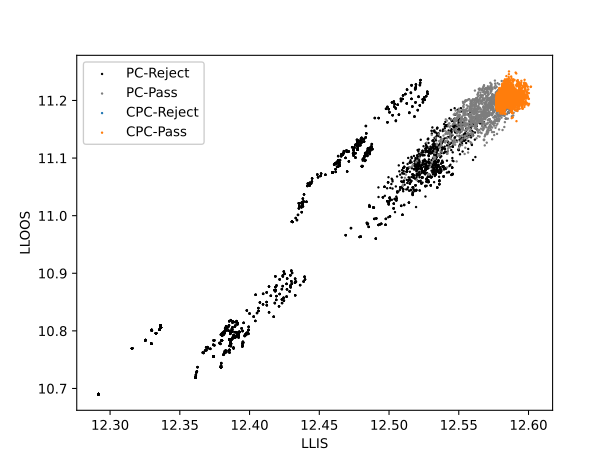}
        \caption{Group1-NoDCC.}
        \label{figure:C4_MCNoDCCISOOSLL}
      \end{minipage} &
      \begin{minipage}[t]{0.5\hsize}
        \centering
        \includegraphics[keepaspectratio, scale=0.55]{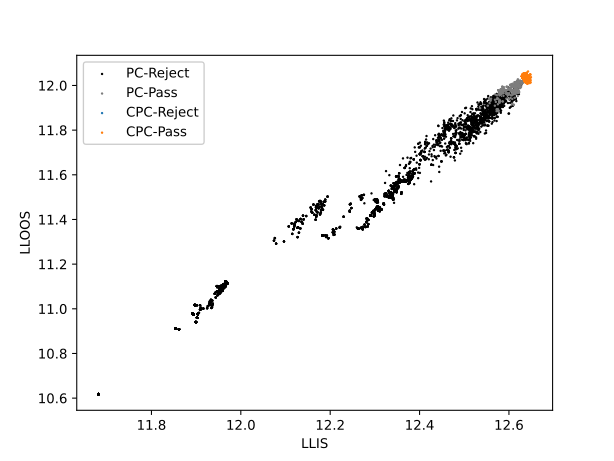}
        \caption{Group2-NoDCC.}
        \label{figure:C4_MCCholeskyISOOSLL}
      \end{minipage}\\
      \begin{minipage}[t]{0.5\hsize}
        \centering
        \includegraphics[keepaspectratio, scale=0.55]{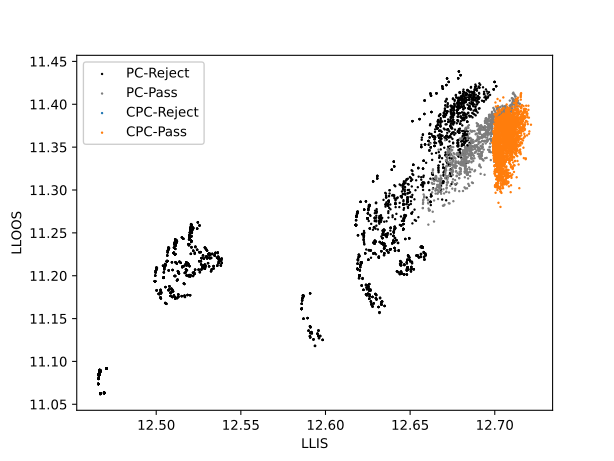}
        \caption{Group3-NoDCC.}
        \label{figure:C4_MCSqrtISOOSLL}
      \end{minipage} &
    \end{tabular}
  \end{minipage}\\
  \begin{minipage}[t]{0.9\hsize}
    {\footnotesize 
    Scatter diagrams plotting LLIS-LLOOS for the estimation results of 5,184 different pair copulas in NoDCC, in which LLIS is on the horizontal axis and LLOOS is on the vertical axis.
    ``Pass'' shows that the points are classified by ``T,'' which means the linear correlations of estimated distribution are included in the confidence intervals reported in Tables \ref{table:C4_Group1CorrConfidentialInterval}-\ref{table:C4_Group3CorrConfidentialInterval}.
    ``Reject'' shows that the points are classified by ``F.''
    Group1 has only 5 points in ``CPC-Reject''; Group2 and Group3 have no ``CPC-Reject'' points.
    }
  \end{minipage}
\end{figure}

Here, we compare the likelihood of log returns for in-sample and out-of-sample data using a combination of correlation matrix decomposition methods and residual distributions.
For each result, the average log likelihoods of the logarithmic returns of the in-sample period (LLIS) and out-of-sample period (LLOOS) are calculated.
Figures \ref{figure:C4_MCNoDCCISOOSLL}-\ref{figure:C4_MCSqrtISOOSLL} plot the LLIS-LLOOS scatter diagrams of NoDCC (CPC) and (PC) settings.
These figures help confirm whether the good performance of the correlation adjustment add-in for the in-sample data is derived from overfitting.
The points are color-coded according to whether they represent the (CPC) or (PC) and whether the linear correlations of the estimated distribution are included in the confidence intervals.
This corresponds to the accepted and rejected points listed in Table \ref{table:C4_CorrelationTestPC}.
In Group1 and Group2, (CPC) is better than (PC) in both the LLIS and LLOOS, indicating that (CPC) improves the likelihood of returns without overfitting.
In contrast, Group3 demonstrates improvement in the LLIS but little improvement in the LLOOS.
Some groups with a better LLOOS exist in (PC), but their correlations are not included in the confidence intervals in Group3.
This indicates that overfitting may occur in low-correlation data in terms of likelihood.

Tables \ref{table:C4_LLGroup1}- \ref{table:C4_LLGroup3} summarize the results of AIC, BIC\footnote{Akaike's information criterion (AIC) and Bayesian information criterion (BIC) are defined as $\mbox{AIC}:= -2L+2k$, $\mbox{BIC}:=-2L+k\log n$, where $L$ is the maximum log likelihood, $k$ is the number of parameters of the model, and $n$ is the number of data.},
and log likelihood for a combination of the decomposition method of the correlation matrix, distribution of residuals, and dataset.
The ``Method'' column shows the decomposition methods for the correlation matrix, and ``Type'' is the type of distribution assumed for the residual.
The ``AIC'' and ``BIC'' columns represent the AIC and BIC values in the estimation.
The ``LLIS'' column is the average log likelihood of the logarithmic returns of the in-sample period.
The ``LLOOS'' column is the average log likelihood of the logarithmic returns of the out-of-sample period.
The ``Use Correlation Adjustment Add-in'' and ``No Correlation Adjustment Add-in'' indicate whether the correlation adjustment add-in is applied or not.
For example, the cells whose Type is ``IC'' and ``Use Correlation Adjustment Add-in'' show the results of (CIC).
The underlined cells show the best-fitting results in the block of the same decomposition method for each column.
The results in bold indicate the highest LLOOS values in each column.
For (PC) and (CPC), the cases with the best AIC, BIC, and LLIS are listed as PC-AIC, PC-BIC, and PC-LLIS, respectively.

\begin{table}[H]
  \centering
  \begin{minipage}[t]{1.0\hsize}
  \centering
  \caption{Fitting results by choice of correlation decomposition and correlation adjustment add-in in Group1}
  \label{table:C4_LLGroup1}
  \begin{tabular}{ll|rrrr|rrrr}
    \hline
             & & \multicolumn{4}{c|}{Use Correlation Adjustment Add-in} & \multicolumn{4}{c}{No Correlation Adjustment Add-in}\\
         Method & Type &  AIC &    BIC &  LLIS &  LLOOS &  AIC &       BIC &    LLIS &   LLOOS \\
    \hline
     NoDCC &         IC & 6144.05 & 6195.36 &  12.5767 &   11.1930 & 6581.06 & 6609.05 & 12.2916 & 10.6902 \\
     NoDCC &         GC & 6143.38 & 6208.68 &  12.5810 &   11.1952 & 6150.96 & 6192.94 & 12.5697 & 11.1975 \\
     NoDCC &         TC & 6126.79 & 6196.76 &  12.5928 &   11.2206 & 6127.60 & 6174.24 & 12.5859 & \underline{11.2187} \\
     NoDCC &    PC-AIC & \underline{6112.75} & \underline{6182.72} &  \underline{12.6018} &   \underline{11.2241} & \underline{6116.29} & \underline{6162.93} & 12.5931 & 11.1857 \\
     NoDCC &    PC-BIC & \underline{6112.75} & \underline{6182.72} &  \underline{12.6018} &   \underline{11.2241} & \underline{6116.29} & \underline{6162.93} & 12.5931 & 11.1857 \\
     NoDCC &   PC-LLIS & \underline{6112.75} & \underline{6182.72} &  \underline{12.6018} &   \underline{11.2241} & 6117.97 & 6169.28 & \underline{12.5933} & 11.1940 \\
     \hline
      Sqrt &         IC & 6565.51 & 6616.81 &  12.5870 &   \underline{\bf{11.2832}} & 6556.36 & \underline{6584.34} & 12.5865 & 11.2715 \\
      Sqrt &         GC & 6560.87 & 6626.17 &  12.5938 &   11.2640 & 6561.96 & 6603.94 & 12.5867 & 11.2745 \\
      Sqrt &         TC & 6551.09 & 6621.05 &  12.6013 &   11.2724 & 6549.08 & 6595.72 & 12.5962 & \underline{\bf{11.2793}} \\
      Sqrt &    PC-AIC & \underline{6535.85} & 6610.48 &  \underline{12.6123} &   11.2600 & \underline{6544.40} & 6595.71 & \underline{12.6005} & 11.2658 \\
      Sqrt &    PC-BIC & 6541.18 & \underline{6606.48} &  12.6063 &   11.2772 & 6547.46 & 6594.10 & 12.5972 & 11.2681 \\
      Sqrt &   PC-LLIS & \underline{6535.85} & 6610.48 &  \underline{12.6123} &   11.2600 & \underline{6544.40} & 6595.71 & \underline{12.6005} & 11.2658 \\
      \hline
     Sqrt2 &         IC & 6567.50 & 6618.81 &  12.5857 &   \underline{11.2786} & 6558.59 & \underline{6586.58} & 12.5850 & 11.2641 \\
     Sqrt2 &         GC & 6558.65 & 6623.96 &  12.5952 &   11.2597 & 6564.09 & 6606.07 & 12.5854 & 11.2678 \\
     Sqrt2 &         TC & 6549.86 & 6619.82 &  12.6021 &   11.2724 & 6550.05 & 6596.70 & 12.5956 & \underline{11.2760} \\
     Sqrt2 &    PC-AIC & \underline{6536.16} & 6610.79 &  \underline{12.6121} &   11.2623 & \underline{6547.37} & 6598.67 & \underline{12.5986} & 11.2604 \\
     Sqrt2 &    PC-BIC & 6540.49 & \underline{6605.79} &  12.6068 &   11.2781 & 6549.68 & 6596.32 & 12.5958 & 11.2622 \\
     Sqrt2 &   PC-LLIS & \underline{6536.16} & 6610.79 &  \underline{12.6121} &   11.2623 & \underline{6547.37} & 6598.67 & \underline{12.5986} & 11.2604 \\
     \hline
  Cholesky &         IC & 6566.55 & 6617.85 &  12.5863 &   11.2462 & 6557.42 & \underline{6585.41} & 12.5858 & 11.2439 \\
  Cholesky &         GC & 6566.34 & 6631.64 &  12.5903 &   11.2525 & 6563.10 & 6605.08 & 12.5860 & 11.2491 \\
  Cholesky &         TC & 6551.64 & 6621.61 &  12.6009 &   11.2730 & 6547.67 & 6594.31 & 12.5971 & \underline{11.2779} \\
  Cholesky &    PC-AIC & \underline{6536.66} & \underline{6606.62} &  \underline{12.6105} &   \underline{11.2775} & \underline{6538.34} & 6589.65 & \underline{12.6043} & 11.2557 \\
  Cholesky &    PC-BIC & \underline{6536.66} & \underline{6606.62} &  \underline{12.6105} &   \underline{11.2775} & 6541.21 & 6587.85 & 12.6012 & 11.2524 \\
  Cholesky &   PC-LLIS & \underline{6536.66} & \underline{6606.62} &  \underline{12.6105} &   \underline{11.2775} & \underline{6538.34} & 6589.65 & \underline{12.6043} & 11.2557 \\
  \hline
     Eigen &         IC & 6601.70 & 6653.01 &  12.5639 &   11.1488 & 6593.24 & 6621.22 & 12.5629 & 11.1568 \\
     Eigen &         GC & 6585.69 & 6651.00 &  12.5780 &   11.1989 & 6598.81 & 6640.79 & 12.5632 & 11.1503 \\
     Eigen &         TC & 6554.84 & 6624.80 &  12.5989 &   11.2391 & \underline{6548.71} & \underline{6595.35} & 12.5964 & \underline{11.2510} \\
     Eigen &    PC-AIC & \underline{6543.37} & 6622.67 &  \underline{12.6088} &   \underline{11.2516} & 6552.09 & 6608.06 & \underline{12.5968} & 11.2497 \\
     Eigen &    PC-BIC & 6544.99 & \underline{6614.96} &  12.6052 &   11.2307 & 6552.09 & 6608.06 & \underline{12.5968} & 11.2497 \\
     Eigen &   PC-LLIS & \underline{6543.37} & 6622.67 &  \underline{12.6088} &   \underline{11.2516} & 6552.09 & 6608.06 & \underline{12.5968} & 11.2497 \\
     \hline
    Eigen2 &         IC & 6599.14 & 6650.45 &  12.5656 &   11.1219 & 6595.54 & 6623.52 & 12.5615 & 11.1501 \\
    Eigen2 &         GC & 6597.48 & 6662.78 &  12.5704 &   11.1204 & 6598.48 & 6640.46 & 12.5634 & 11.1232 \\
    Eigen2 &         TC & 6549.67 & \underline{6619.63} &  12.6022 &   11.2393 & \underline{6547.25} & \underline{6593.89} & 12.5974 & \underline{11.2324} \\
    Eigen2 &    PC-AIC & \underline{6545.78} & 6620.41 &  \underline{12.6060} &   \underline{11.2572} & 6549.82 & 6605.79 & \underline{12.5983} & 11.2274 \\
    Eigen2 &    PC-BIC & 6550.11 & 6620.08 &  12.6019 &   11.2139 & 6549.82 & 6605.79 & \underline{12.5983} & 11.2274 \\
    Eigen2 &   PC-LLIS & \underline{6545.78} & 6620.41 &  \underline{12.6060} &   \underline{11.2572} & 6549.82 & 6605.79 & \underline{12.5983} & 11.2274 \\
    \hline
    \end{tabular}
  \end{minipage}
  \begin{minipage}[t]{0.9\hsize}
    {\footnotesize The table summarizes the AIC, BIC, and log-likelihood results for the estimation of the model in Group1.
The ``Method'' columns represent the decomposition methods for the correlation matrix; The ``Type'' is the type of distribution to be fitted to the residual.
The ``AIC'' and ``BIC'' columns represent the AIC and BIC values in the estimation.
The ``LLIS'' columns are the average log likelihood of the logarithmic returns of the in-sample period.
The ``LLOOS'' columns are the average log likelihood of the logarithmic returns of the out-of-sample period.
The ``Use Correlation Adjustment Add-in'' and ``No Correlation Adjustment Add-in'' indicate whether the correlation adjustment add-in is applied or not.
The underlined cells show the best-fitting results in the block using the same decomposition method for each column.
The results in bold indicate the highest LLOOS values in each column.
For (PC) and (CPC), the cases with the best AIC, BIC, and LLIS are listed as PC-AIC, PC-BIC, and PC-LLIS, respectively.}
  \end{minipage}
\end{table}
  
  \begin{table}[H]
  \centering
  \begin{minipage}[t]{1.0\hsize}
  \centering
  \caption{Fitting results by choice of correlation decomposition and correlation adjustment add-in in Group2}
  \label{table:C4_LLGroup2}
  \begin{tabular}{ll|lrrr|lrrr}
    \hline
    & & \multicolumn{4}{c|}{Use Correlation Adjustment Add-in} & \multicolumn{4}{c}{No Correlation Adjustment Add-in}\\
    Method & Type &  AIC &    BIC &  LLIS &  LLOOS &  AIC &       BIC &    LLIS &   LLOOS \\
    \hline
     NoDCC &         IC & 5101.24 & 5152.55 &  12.6294 &   \underline{12.0410} & 6577.89 & 6605.88 & 11.6812 & 10.6177 \\
     NoDCC &         GC & 5096.24 & 5161.54 &  12.6364 &   12.0291 & 5112.05 & 5154.03 & 12.6199 & 12.0015 \\
     NoDCC &         TC & 5093.03 & 5163.00 &  12.6397 &   12.0356 & 5092.95 & \underline{5139.60} & 12.6334 & \underline{12.0294} \\
     NoDCC &    PC-AIC & \underline{5077.97} & \underline{5147.93} &  \underline{12.6493} &   12.0370 & \underline{5090.01} & 5141.32 & \underline{12.6365} & 12.0267 \\
     NoDCC &    PC-BIC & \underline{5077.97} & \underline{5147.93} &  \underline{12.6493} &   12.0370 & \underline{5090.01} & 5141.32 & \underline{12.6365} & 12.0267 \\
     NoDCC &   PC-LLIS & \underline{5077.97} & \underline{5147.93} &  \underline{12.6493} &   12.0370 & \underline{5090.01} & 5141.32 & \underline{12.6365} & 12.0267 \\
     \hline
      Sqrt &         IC & 6595.04 & \underline{6646.35} &  12.6503 &   12.0763 & 6586.00 & \underline{6613.98} & 12.6497 & 12.0788 \\
      Sqrt &         GC & 6591.59 & 6656.89 &  12.6564 &   12.0786 & 6591.61 & 6633.59 & 12.6500 & 12.0757 \\
      Sqrt &         TC & 6591.61 & 6661.57 &  12.6576 &   12.0781 & 6587.17 & 6633.81 & 12.6541 & 12.0787 \\
      Sqrt &    PC-AIC & \underline{6579.97} & 6649.94 &  \underline{12.6650} &   \underline{12.0809} & \underline{6585.84} & 6627.82 & 12.6536 & 12.0788 \\
      Sqrt &    PC-BIC & \underline{6579.97} & 6649.94 &  \underline{12.6650} &   \underline{12.0809} & \underline{6585.84} & 6627.82 & 12.6536 & 12.0788 \\
      Sqrt &   PC-LLIS & \underline{6579.97} & 6649.94 &  \underline{12.6650} &   \underline{12.0809} & 6588.10 & 6639.40 & \underline{12.6548} & \underline{12.0797} \\
      \hline
     Sqrt2 &         IC & 6596.95 & \underline{6648.26} &  12.6491 &   12.0820 & 6587.70 & \underline{6615.69} & 12.6486 & 12.0811 \\
     Sqrt2 &         GC & 6594.83 & 6660.14 &  12.6543 &   12.0864 & 6593.49 & 6635.47 & 12.6488 & 12.0813 \\
     Sqrt2 &         TC & 6592.10 & 6662.07 &  12.6573 &   12.0817 & 6588.37 & 6635.01 & 12.6533 & 12.0833 \\
     Sqrt2 &    PC-AIC & \underline{6581.60} & 6651.57 &  \underline{12.6640} &   \underline{12.0866} & \underline{6587.18} & 6629.16 & 12.6528 & \underline{12.0846} \\
     Sqrt2 &    PC-BIC & \underline{6581.60} & 6651.57 &  \underline{12.6640} &   \underline{12.0866} & \underline{6587.18} & 6629.16 & 12.6528 & \underline{12.0846} \\
     Sqrt2 &   PC-LLIS & \underline{6581.60} & 6651.57 &  \underline{12.6640} &   \underline{12.0866} & 6591.31 & 6647.28 & \underline{12.6540} & 12.0823 \\
     \hline
  Cholesky &         IC & 6597.21 & 6648.52 &  12.6489 &   \underline{12.098214} & 6587.68 & \underline{6615.67} & 12.6486 & 12.0992 \\
  Cholesky &         GC & 6595.33 & 6660.63 &  12.6540 &   12.098211 & 6593.44 & 6635.42 & 12.6488 & 12.1009 \\
  Cholesky &         TC & 6584.32 & 6654.28 &  12.6623 &   12.0941 & 6585.89 & 6632.53 & 12.6549 & 12.0946 \\
  Cholesky &    PC-AIC & \underline{6580.62} & \underline{6645.92} &  12.6633 &   12.0736 & \underline{6580.19} & 6626.83 & \underline{12.6585} & \underline{12.1012} \\
  Cholesky &    PC-BIC & \underline{6580.62} & \underline{6645.92} &  12.6633 &   12.0736 & \underline{6580.19} & 6626.83 & \underline{12.6585} & \underline{12.1012} \\
  Cholesky &   PC-LLIS & 6583.33 & 6662.63 &  \underline{12.6654} &   12.0805 & \underline{6580.19} & 6626.83 & \underline{12.6585} & \underline{12.1012} \\
  \hline
     Eigen &         IC & 6621.01 & 6672.32 &  12.6338 &   \underline{\bf{12.1060}} & 6611.92 & 6639.91 & 12.6332 & \underline{\bf{12.1075}} \\
     Eigen &         GC & 6599.47 & 6664.77 &  12.6513 &   12.0859 & 6617.24 & 6659.22 & 12.6336 & 12.1062 \\
     Eigen &         TC & 6591.23 & 6661.19 &  12.6579 &   12.0812 & \underline{6587.95} & \underline{6634.60} & 12.6536 & 12.0819 \\
     Eigen &    PC-AIC & \underline{6580.89} & 6655.52 &  \underline{12.6657} &   12.0781 & 6590.08 & 6646.05 & \underline{12.6548} & 12.0769 \\
     Eigen &    PC-BIC & 6583.94 & \underline{6653.90} &  12.6625 &   12.0705 & 6591.31 & 6642.62 & 12.6527 & 12.0827 \\
     Eigen &   PC-LLIS & \underline{6580.89} & 6655.52 &  \underline{12.6657} &   12.0781 & 6590.08 & 6646.05 & \underline{12.6548} & 12.0769 \\
     \hline
    Eigen2 &         IC & 6614.10 & 6665.41 &  12.6382 &   \underline{12.1013} & 6606.63 & 6634.62 & 12.6366 & \underline{12.1050} \\
    Eigen2 &         GC & 6604.00 & 6669.30 &  12.6484 &   12.0794 & 6610.85 & 6652.83 & 12.6377 & 12.1037 \\
    Eigen2 &         TC & 6583.15 & 6653.12 &  12.6630 &   12.0789 & \underline{6582.97} & \underline{6629.62} & 12.6567 & 12.0839 \\
    Eigen2 &    PC-AIC & \underline{6578.59} & 6653.22 &  \underline{12.6672} &   12.0801 & 6585.50 & 6641.47 & \underline{12.6577} & 12.0807 \\
    Eigen2 &    PC-BIC & 6582.81 & \underline{6652.77} &  12.6632 &   12.0811 & 6586.91 & 6638.22 & 12.6555 & 12.0865 \\
    Eigen2 &   PC-LLIS & \underline{6578.59} & 6653.22 &  \underline{12.6672} &   12.0801 & 6585.50 & 6641.47 & \underline{12.6577} & 12.0807 \\
    \hline
  \end{tabular}
  \end{minipage}
  \begin{minipage}[t]{0.9\hsize}
    {\footnotesize 
The table summarizes the AIC, BIC, and log likelihood calculated from the estimation results for the tested models in Group2.
See Table \ref{table:C4_LLGroup1} for the explanation of each row and column.}
  \end{minipage}
  \end{table}
    
  \begin{table}[H]
  \centering
  \begin{minipage}[t]{1.0\hsize}
  \centering
  \caption{Fitting results by choice of correlation decomposition and correlation adjustment add-in in Group3}
  \label{table:C4_LLGroup3}
  \begin{tabular}{ll|lrrr|lrrr}
    \hline
    & & \multicolumn{4}{c|}{Use Correlation Adjustment Add-in} & \multicolumn{4}{c}{No Correlation Adjustment Add-in}\\
    Method & Type &  AIC &    BIC &  LLIS &  LLOOS &  AIC &       BIC &    LLIS &   LLOOS \\
    \hline
     NoDCC &         IC & 6206.58 & 6257.89 &  12.6990 &   11.3546 & 6562.22 & 6590.20 & 12.4658 & 11.0737 \\
     NoDCC &         GC & 6205.11 & 6270.41 &  12.7038 &   11.3113 & 6215.38 & 6257.36 & 12.6909 & 11.3691 \\
     NoDCC &         TC & 6184.86 & 6254.82 &  12.7180 &   11.3726 & 6183.12 & 6229.76 & 12.7127 & 11.4011 \\
     NoDCC &    PC-AIC & 6181.87 & 6251.84 &  12.7199 &   \underline{11.3983} & \underline{6181.74} & \underline{6228.38} & 12.7136 & \underline{11.4013} \\
     NoDCC &    PC-BIC & 6184.63 & \underline{6249.94} &  12.7168 &   11.3760 & \underline{6181.74} & \underline{6228.38} & 12.7136 & \underline{11.4013} \\
     NoDCC &   PC-LLIS & \underline{6183.51} & 6262.81 &  \underline{12.7214} &   11.3761 & 6181.75 & 6233.05 & \underline{12.7149} & 11.4005 \\
     \hline
      Sqrt &         IC & 6580.88 & 6632.18 &  12.7149 &   11.4418 & 6571.48 & \underline{6599.47} & 12.7145 & 11.4310 \\
      Sqrt &         GC & 6585.12 & 6650.42 &  12.7160 &   \underline{\bf{11.4434}} & 6577.18 & 6619.16 & 12.7147 & \underline{\bf{11.4370}} \\
      Sqrt &         TC & 6556.59 & 6626.56 &  12.7355 &   11.4020 & \underline{6555.96} & 6602.61 & 12.7295 & 11.4363 \\
      Sqrt &    PC-AIC & \underline{6553.04} & \underline{6623.01} &  12.7377 &   11.4157 & 6557.00 & 6608.31 & 12.7301 & 11.4217 \\
      Sqrt &    PC-BIC & \underline{6553.04} & \underline{6623.01} &  12.7377 &   11.4157 & 6559.31 & 6605.96 & 12.7274 & 11.4163 \\
      Sqrt &   PC-LLIS & 6554.29 & 6628.92 &  \underline{12.7382} &   11.4167 & 6558.00 & 6613.97 & \underline{12.7308} & 11.4276 \\
     \hline
     Sqrt2 &         IC & 6581.26 & 6632.57 &  12.7146 &   11.4381 & 6571.89 & \underline{6599.88} & 12.7142 & 11.4281 \\
     Sqrt2 &         GC & 6585.91 & 6651.21 &  12.7155 &   \underline{11.4411} & 6577.55 & 6619.53 & 12.7145 & 11.4345 \\
     Sqrt2 &         TC & 6566.23 & 6636.20 &  12.7293 &   11.4373 & 6556.34 & 6602.98 & 12.7293 & \underline{11.4360} \\
     Sqrt2 &    PC-AIC & \underline{6552.01} & \underline{6621.98} &  \underline{12.7384} &   11.4142 & \underline{6555.97} & 6607.28 & 12.7308 & 11.4197 \\
     Sqrt2 &    PC-BIC & \underline{6552.01} & \underline{6621.98} &  \underline{12.7384} &   11.4142 & 6557.77 & 6604.41 & 12.7283 & 11.4143 \\
     Sqrt2 &   PC-LLIS & 6553.96 & 6628.59 &  \underline{12.7384} &   11.4169 & 6557.61 & 6613.59 & \underline{12.7310} & 11.4252 \\
     \hline
  Cholesky &         IC & 6577.41 & 6628.72 &  12.7171 &   11.3902 & 6568.36 & \underline{6596.35} & 12.7165 & 11.3883 \\
  Cholesky &         GC & 6578.55 & 6643.85 &  12.7202 &   11.3487 & 6573.91 & 6615.89 & 12.7168 & 11.3908 \\
  Cholesky &         TC & 6557.62 & 6627.59 &  12.7348 &   11.4028 & \underline{6556.06} & 6602.70 & 12.7294 & \underline{11.4183} \\
  Cholesky &    PC-AIC & \underline{6554.36} & 6629.00 &  \underline{12.7382} &   11.4012 & 6557.68 & 6604.32 & 12.7284 & 11.3882 \\
  Cholesky &    PC-BIC & 6559.31 & \underline{6624.61} &  12.7325 &   \underline{11.4055} & 6560.82 & 6602.80 & 12.7251 & 11.3710 \\
  Cholesky &   PC-LLIS & \underline{6554.36} & 6629.00 &  \underline{12.7382} &   11.4012 & 6559.01 & 6614.98 & \underline{12.7301} & 11.4175 \\
     \hline
     Eigen &         IC & 6603.28 & 6654.58 &  12.7006 &   11.2948 & 6593.46 & 6621.45 & 12.7005 & 11.2966 \\
     Eigen &         GC & 6599.44 & 6664.74 &  12.7069 &   11.3319 & 6599.29 & 6641.27 & 12.7006 & 11.2978 \\
     Eigen &         TC & \underline{6547.63} & \underline{6617.60} &  \underline{12.7412} &   11.4069 & \underline{6554.73} & \underline{6601.38} & 12.7303 & 11.4038 \\
     Eigen &    PC-AIC & 6550.60 & 6625.23 &  12.7406 &   11.3981 & 6558.34 & 6614.31 & \underline{12.7305} & \underline{11.4061} \\
     Eigen &    PC-BIC & 6556.79 & 6622.09 &  12.7341 &   11.3656 & 6561.38 & 6612.69 & 12.7273 & 11.4026 \\
     Eigen &   PC-LLIS & 6552.26 & 6631.56 &  12.7408 &   \underline{11.4079} & 6558.34 & 6614.31 & \underline{12.7305} & \underline{11.4061} \\
     \hline
    Eigen2 &         IC & 6612.30 & 6663.61 &  12.6948 &   11.3102 & 6604.68 & 6632.67 & 12.6933 & 11.3176 \\
    Eigen2 &         GC & 6599.80 & 6665.10 &  12.7066 &   11.3444 & 6609.17 & 6651.15 & 12.6943 & 11.3021 \\
    Eigen2 &         TC & \underline{6553.57} & \underline{6623.53} &  12.7374 &   11.3821 & \underline{6557.13} & \underline{6603.78} & 12.7288 & 11.3879 \\
    Eigen2 &    PC-AIC & 6554.94 & 6634.23 &  \underline{12.7391} &   11.3949 & 6559.72 & 6615.69 & \underline{12.7297} & \underline{11.3906} \\
    Eigen2 &    PC-BIC & 6559.14 & 6629.10 &  12.7338 &   \underline{11.4033} & 6561.48 & 6612.79 & 12.7273 & 11.3870 \\
    Eigen2 &   PC-LLIS & 6554.94 & 6634.23 &  \underline{12.7391} &   11.3949 & 6559.72 & 6615.69 & \underline{12.7297} & \underline{11.3906} \\
    \hline
    \end{tabular}
  \end{minipage}
  \begin{minipage}[t]{0.9\hsize}
    {\footnotesize 
The table summarizes the AIC, BIC, and log likelihood calculated from the estimation results for the tested models in Group3.
See Table \ref{table:C4_LLGroup1} for the explanation of each row and column.}
  \end{minipage}
\end{table}

The LLIS and LLOOS of Sqrt and Sqrt2 outperform the LLIS and LLOOS of NoDCC in all the data groups.
This result indicates that the DCC-GARCH model with a Sqrt or Sqrt2 decomposition is better than the copula-GARCH model.
In addition, the LLIS of the pair copula is the largest except the block \{Group3, Eigen, ``Use Correlation Adjustment Add-in''\}.
In other words, fitting the in-sample data is the best when the pair copula is used in most cases.
However, for the LLOOS, there are cases in which the best result for each block is obtained by the parametric copula, and the best results for each asset group are obtained using a parametric copula.
This indicated that the pair copula tends to overfit and even a simple copula model can provide a good out-of-sample fitting if the decomposition method is appropriately chosen.
By comparing the correlation matrix decomposition methods in DCC-GARCH,
we observe that Sqrt and Sqrt2 are better than Colesky, whereas Eigen and Eigen2 are worse.
Other studies show that Sqrt and Sqrt2 are worse than Colesky, whereas Eigen and Eigen2 are better.
The order did not change depending on whether a correlation adjustment add-in is applied.
Cholesky's performance is stable, regardless of the level of correlation.
LLOOS comparisons of Sqrt and Sqrt2 and Eigen and Eigen2 show that the results for Sqrt and Eigen are better in many cases.
This suggests that a formulation that decomposes the correlation matrix is more effective.
The maximum LLOOS is observed in Group1 when (CIC) is used for Sqrt decomposition, in Group2 when (IC) is used for Eigen decomposition, and in Group3 when (GCG) is used for Sqrt decomposition.
For the (CGC), the Gaussian copula parameters are estimated as:
\begin{eqnarray}
  \Sigma_{G}=\begin{pmatrix}1 & 0.00606 & 0.00006\\ -0.00606 & 1 & 0.01556\\0.00006 & 0.01556 &1\end{pmatrix}
\end{eqnarray}
that is close to that of an independent copula.
This shows that the correlation structure is simple and almost independent under appropriate settings, thanks to the decomposition methods and correlation adjustment add-in.

\begin{figure}[H]
\centering
\begin{minipage}[t]{0.9\hsize}
  \centering
  \begin{tabular}{cc}
    \begin{minipage}[t]{0.5\hsize}
      \centering
      \includegraphics[keepaspectratio, scale=0.55]{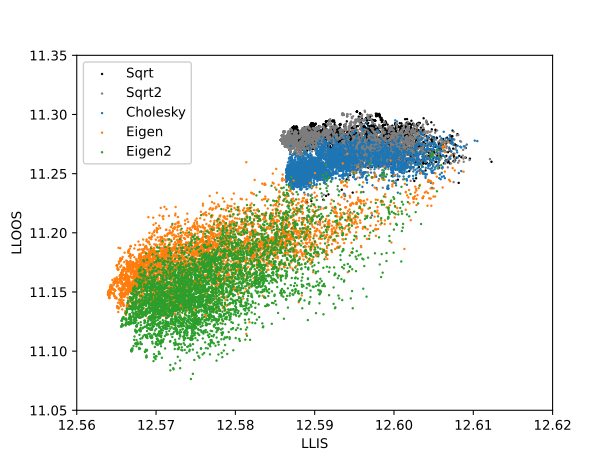}
      \caption{LLIS-LLOOS plot by group1 and (CPC). }
      \label{figure:C4_MCCLLPlot}
    \end{minipage} &
    \begin{minipage}[t]{0.5\hsize}
      \centering
      \includegraphics[keepaspectratio, scale=0.55]{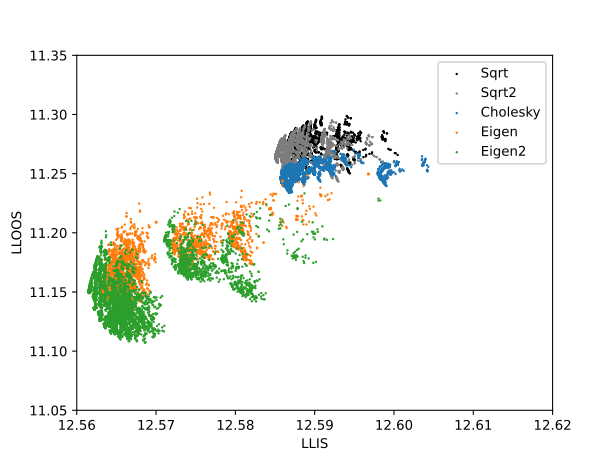}
      \caption{LLIS-LLOOS plot by group1 and (PC). }
      \label{figure:C4_MCNLLPlot}
    \end{minipage}\\
    \begin{minipage}[t]{0.5\hsize}
      \centering
      \includegraphics[keepaspectratio, scale=0.55]{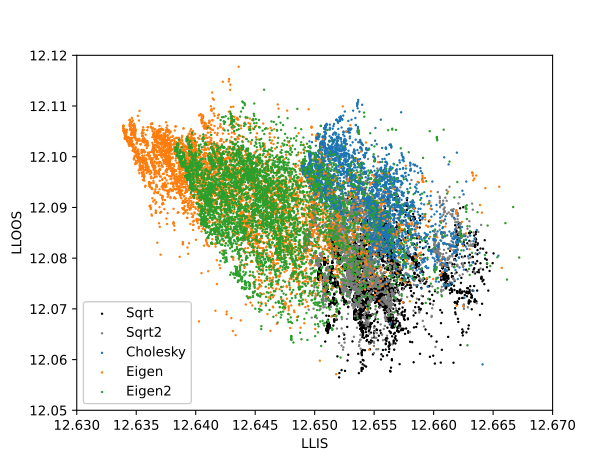}
      \caption{LLIS-LLOOS plot by group2 and (CPC). }
      \label{figure:C4_RCCLLPlot}
    \end{minipage} &
    \begin{minipage}[t]{0.5\hsize}
      \centering
      \includegraphics[keepaspectratio, scale=0.55]{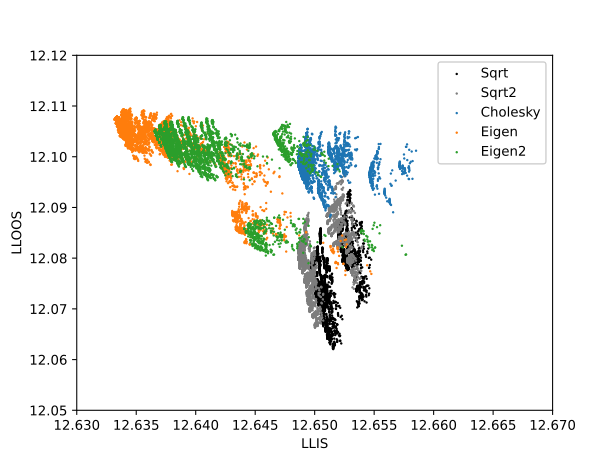}
      \caption{LLIS-LLOOS plot by group2 and (PC). }
      \label{figure:C4_RCNLLPlot}
    \end{minipage}\\
    \begin{minipage}[t]{0.5\hsize}
      \centering
      \includegraphics[keepaspectratio, scale=0.55]{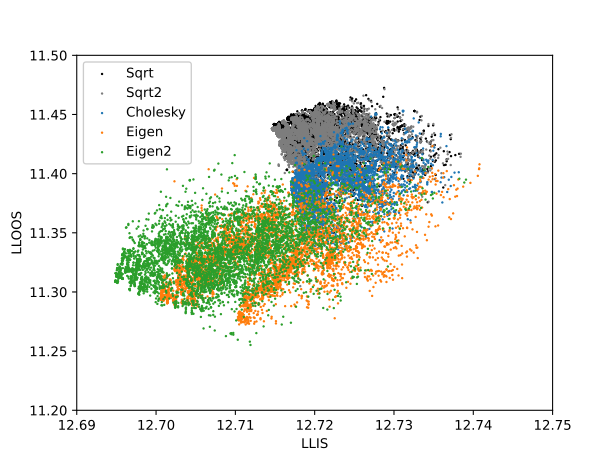}
      \caption{LLIS-LLOOS plot by group3 and (CPC). }
      \label{figure:C4_NCCLLPlot}
    \end{minipage} &
    \begin{minipage}[t]{0.5\hsize}
      \centering
      \includegraphics[keepaspectratio, scale=0.55]{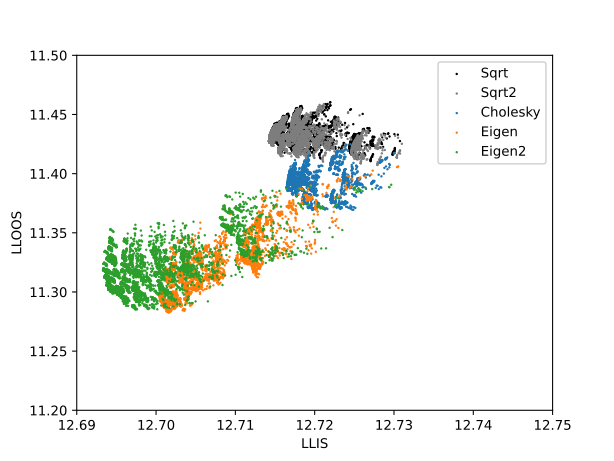}
      \caption{LLIS-LLOOS plot by group2 and (PC). }
      \label{figure:C4_NCNLLPlot}
    \end{minipage}
  \end{tabular}
\end{minipage}
\\
\begin{minipage}[t]{0.9\hsize}
  {\footnotesize
The LLIS-LLOOS plots computed from the results of fitting the DCC residual to the distributions (PC) and (CPC) using the pair copula.
One point corresponds to an estimated result by 1 of the 5,184 different pair copulas.
Sqrt, Sqrt2, Cholesky, Eigen, and Eigen2 in the legend are the decomposition methods of the correlation matrix.}
\end{minipage}
\end{figure}

Figures \ref{figure:C4_MCCLLPlot}-\ref{figure:C4_NCNLLPlot} show the LLIS-LLOOS of (CPC) and (PC) estimation results according to the decomposition method.
In Group1 and Group3, LLOOS levels show that Sqrt and Sqrt2 are better than Cholesky, whereas Eigen and Eigen2 are worse than Cholesky.
However, in Group2, Sqrt and Sqrt2 are worse than Cholesky, and Eigen and Eigen2 are better.
This finding is consistent with the results in Tables \ref{table:C4_LLGroup1}-\ref{table:C4_LLGroup3}.
Under these settings, Eigen and Eigen2 exhibit the best performance in LLOOS fitting for data with a high correlation.
On the LLIS axis, Sqrt, Sqrt2, and Cholesky are located at similar levels and tend to be better than Eigen and Eigen2.
When comparing the correlation and covariance matrix decompositions,
Sqrt and Sqrt2 had similar shapes in almost all figures, whereas Eigen and Eigen2 show similar trends.
In Group1 and Group3, Sqrt and Eigen tend to be better than Sqrt2 and Eigen2, respectively, although the differences are small.
In Group2, Sqrt2 and Eigen2 are better distributed in LLIS, whereas Sqrt and Eigen are better distributed in LLOOS.

Thus, using the eigenvalue decomposition method is recommended when the correlation is high, and the square root of the matrix is better when the correlation is not high in our example.

\section{Conclusion}
This study investigates the methodologies used to filter out correlations from multivariate time series random variables.
First, we develop a correlation adjustment add-in to adjust the correlation of random variables with parameterized linear transformation.
The transformation forms a lower triangular matrix, and the correlations are estimated using the maximum likelihood estimation.
This method has an advantage over Lee and Long's method because it does not require the computation of the covariance of the original distribution
and can be applied to construct a non-zero correlation distribution.
Second, we use a new matrix decomposition method to convert the GARCH residual into the DCC residual using eigenvalue decomposition.
It takes advantage of the angle of eigenvectors to ensure the consistent decomposition over time.
The decomposition also varies depending on whether it is based on a correlation or covariance matrix.
We compare the performances among the square root of the matrix, Cholesky decomposition, and eigenvalue decomposition with the correlation and covariance matrices.

In the empirical analysis using foreign exchange rates, the correlation adjustment add-in achieves high out-of-sample performance in modeling the residuals in the copula-GARCH model.
This is especially true for data with high levels of correlation owing to the add-in's ability to fit the correlation of the distribution to the data.
The comparison of decomposition variations shows that the square root of the matrix or the eigenvalue decomposition performs best in the out-of-sample fitting.
When the correlations in the rates of return are large in absolute value, eigenvalue decomposition demonstrates the best performance.
The choice between the correlation and covariance matrices indicates that the former has a slight advantage in terms of out-of-sample fitting. 
In addition to applying the correlation adjustment add-in method,
when the decomposition of the correlation matrix is appropriately chosen,
the best out-of-sample performance is obtained when almost all independent copulas are used.
This suggests that the DCC-GARCH residual can be independent without assuming a complex correlation structure.

\printbibliography

@article{drehmann2023papers,
  title={Global tightening, banking stress and market resilience in EMEs},
  author={Drehmann, Mathias and Maronoti, Bafundi and O’Connor, Angela},
  institution = {Bank for International Settlements},
  type = {BIS Papers},
  number = {134},
  year={2023}
}

@article{mazur2021covid,
  title={COVID-19 and the march 2020 stock market crash. Evidence from S\&P1500},
  author={Mazur, Mieszko and Dang, Man and Vega, Miguel},
  journal={Finance research letters},
  volume={38},
  pages={101690},
  year={2021},
  publisher={Elsevier}
}

@article{pelagatti2004dynamic,
  title={Dynamic conditional correlation with elliptical distributions},
  author={Pelagatti, Matteo M},
  journal={Available at SSRN 888732},
  year={2004}
}

@article{kim2016linear,
  title={Linear time-varying regression with Copula--DCC--GARCH models for volatility},
  author={Kim, Jong-Min and Jung, Hojin},
  journal={Economics Letters},
  volume={145},
  pages={262--265},
  year={2016},
  publisher={Elsevier}
}

@article{ramchand1998volatility,
  title={Volatility and cross correlation across major stock markets},
  author={Ramchand, Latha and Susmel, Raul},
  journal={Journal of Empirical Finance},
  volume={5},
  number={4},
  pages={397--416},
  year={1998},
  publisher={Elsevier}
}

@article{bauwens2006multivariate,
  title={Multivariate GARCH models: a survey},
  author={Bauwens, Luc and Laurent, S{\'e}bastien and Rombouts, Jeroen VK},
  journal={Journal of applied econometrics},
  volume={21},
  number={1},
  pages={79--109},
  year={2006},
  publisher={Wiley Online Library}
}

@article{engle2002dynamic,
  title={Dynamic conditional correlation: A simple class of multivariate generalized autoregressive conditional heteroskedasticity models},
  author={Engle, Robert},
  journal={Journal of Business \& Economic Statistics},
  volume={20},
  number={3},
  pages={339--350},
  year={2002},
  publisher={Taylor \& Francis}
}

@article{aloui2016relationship,
  title={Relationship between oil, stock prices and exchange rates: A vine copula based GARCH method},
  author={Aloui, Riadh and A{\"\i}ssa, Mohamed Safouane Ben},
  journal={The North American Journal of Economics and Finance},
  volume={37},
  pages={458--471},
  year={2016},
  publisher={Elsevier}
}

@article{lee2009copula,
  title={Copula-based multivariate GARCH model with uncorrelated dependent errors},
  author={Lee, Tae-Hwy and Long, Xiangdong},
  journal={Journal of Econometrics},
  volume={150},
  number={2},
  pages={207--218},
  year={2009},
  publisher={Elsevier}
}

@article{kimani2023modelling,
  title={Modelling Dependence of Cryptocurrencies Using Copula Garch},
  author={Kimani, Eric M and Ngunyi, Anthony and Mungatu, Joseph K},
  journal={Journal of Mathematical Finance},
  volume={13},
  number={3},
  pages={321--338},
  year={2023},
  publisher={Scientific Research Publishing}
}

@book{jaworski2010copula,
  title={Copula theory and its applications},
  author={Jaworski, Piotr and Durante, Fabrizio and Hardle, Wolfgang Karl and Rychlik, Tomasz},
  volume={198},
  year={2010},
  publisher={Springer}
}

@article{wurtz2006parameter,
  title={Parameter estimation of ARMA models with GARCH/APARCH errors an R and SPlus software implementation},
  author={Wurtz, Diethelm and Chalabi, Yohan and Luksan, Ladislav},
  journal={Journal of Statistical Software},
  volume={55},
  number={2},
  pages={28--33},
  year={2006}
}

@article{engle1995multivariate,
  title={Multivariate simultaneous generalized ARCH},
  author={Engle, Robert and Kroner, Kenneth},
  journal={Econometric theory},
  volume={11},
  number={1},
  pages={122--150},
  year={1995},
  publisher={Cambridge University Press}
}

@techreport{troben2007,
author = {Troben, G. Andersen and Tim, Bollerslev and Peter, Christoffersen and Francis, X. Diebold},
institution = {National Bureau of Economic Research, Inc.},
title = {Practical volatility and correlation modeling for financial market risk management},
type = {NBER Working Papers},
number = {11069},
year = {2007}
}

\end{document}